\documentclass[preprint,authoryear,3p,10pt]{elsarticle}

\makeatletter
\def\ps@pprintTitle{%
 \let\@oddhead\@empty
 \let\@evenhead\@empty
\def\@oddfoot{\centerline{\thepage}}%
 \let\@evenfoot\@oddfoot}
\makeatother

\usepackage{setspace}
\usepackage{graphics}
\usepackage{array,multirow,graphics}

\usepackage{amssymb,amsmath}
\usepackage{amsthm}

\usepackage{booktabs}

\usepackage{hyperref}
\usepackage{cleveref}

\usepackage{dsfont} 

\PassOptionsToPackage{hyphens}{url}\usepackage{hyperref}

\makeatletter
\g@addto@macro{\UrlBreaks}{\UrlOrds}
\makeatother

\usepackage{color}
\definecolor{schalkeblau}{RGB}{0,0,204}
\definecolor{schwarz}{RGB}{0,0,0}
\newcommand{\blue}[1]{\textcolor{schwarz}{#1}}
 \newcommand{\blau}[1]{\textcolor{schwarz}{#1}}

\newtheorem{remark}{Remark}[section]
\newtheorem{lemma}{Lemma}[section]

\newtheorem{definition}{Definition}[section]
\newtheorem{example}{Example}[section] 

\journal{Computers and Operations Research}

\begin{document}

\begin{frontmatter}



\title{Dealing with the Dimensionality Curse in Dynamic Pricing Competition: 
Using Frequent Repricing to Compensate Imperfect Market Anticipations}

\tnotetext[mytitlenote]{Accepted Manuscript. \textcopyright \enskip 2018. This manuscript version is made available under the CC-BY-NC-ND 4.0 license \mbox{\url{http://creativecommons.org/licenses/by-nc-nd/4.0/}}.\\Final version published in Computers \& Operations Research, Volume 100: \url{https://doi.org/10.1016/j.cor.2018.07.011}.
\hspace{5mm}}


\author{Rainer Schlosser}
\ead{rainer.schlosser@hpi.de}
\author{Martin Boissier}
\ead{martin.boissier@hpi.de}
\address{Hasso Plattner Institute, University of Potsdam, Potsdam, Germany}

\begin{abstract}
Most sales applications are characterized by competition and limited demand information. For successful pricing strategies, frequent price adjustments as well as anticipation of market dynamics are crucial. Both effects are challenging as competitive markets are complex and computations of optimized pricing adjustments can be time-consuming. We analyze stochastic dynamic pricing models under oligopoly competition for the sale of perishable goods. To circumvent the curse of dimensionality, we propose a heuristic approach to efficiently compute price adjustments. To demonstrate our strategy's applicability even if the number of competitors is large and their strategies are unknown, we consider different competitive settings in which competitors frequently and strategically adjust their prices. For all settings, we verify that our heuristic strategy yields promising results. We compare the performance of our heuristic against upper bounds, which are obtained by optimal strategies that take advantage of perfect price anticipations. We find that price adjustment frequencies can have a larger impact on expected profits than price anticipations. 
Finally, our approach has been applied on Amazon for the sale of used books. We have used a seller's historical market data to calibrate our model. Sales results show that our data-driven strategy outperforms the rule-based strategy of an experienced seller by a profit increase of more than 20\%.
\end{abstract}


\end{frontmatter}


\section{Introduction}

In most markets, sellers have to deal with competition. On modern e-commerce platforms, it has become easy to observe and to adjust prices. On such platforms, sales do not only depend on a sellers' prices but also on competitors' prices. Moreover, customers take further product features into account, e.g., product quality, ratings, shipping time, shipping costs, etc. 

\subsection{Motivation}


Modern market platforms such as Amazon Marketplace or eBay are highly dynamic as sellers can observe the current market situation at any time and adjust their prices instantly. 
Sellers typically rival against dozens of competitors, decide on prices for thousands of products, and face steadily changing market situations, cf. Profitero (2014). 
%
As electronic markets are constantly growing and the costs of price changes are close to free, intelligent automated repricing systems become increasingly important for practitioners, see Popescu (2016).
Since demand information is limited and market dynamics are uncertain, deriving intelligent pricing strategies is highly complex.
Pricing strategies are required to take a multitude of dimensions for each competitor into account (e.g., price, quality, shipping time, shipping costs, ratings). Further, aspects such as discounting, cash-flows, and inventory holding costs also have to be included.

In this paper, we study oligopoly pricing models in a stochastic dynamic finite horizon framework. We consider single-product problems. We seek to deal with the following real-life aspects: 

(i) \hspace{0.3cm}demand probabilities are unknown,

(ii) \hspace{0.20cm}products have multiple offer dimensions (price, quality, ratings, shipping time, etc.),

(iii) \hspace{0.11cm}products are offered by multiple competitors,

(iv) \hspace{0.11cm}competitors' sales, inventories, and strategies are not observable. 
 
\medskip

The setting of limited and asymmetric information, multiple offer dimensions and multiple competitors is highly challenging.
%
To compute viable pricing strategies in competitive settings, we introduce a dynamic programming model that circumvents the curse of dimensionality. We show how to calibrate the model using data-driven demand estimations. 
To illustrate the applicability of our approach, we use reproducible numerical examples.
Finally, we present the performance result of our strategy applied on Amazon. 

\subsection{Literature Review}

The best way to sell products is employing a classical application of revenue management theory. The problem is closely related to the field of dynamic pricing, which is summarized in the books by Talluri, van Ryzin (2004), Phillips (2005), and Yeoman, McMahon-Beattie (2011). The surveys by Chen, Chen (2015) and Den Boer (2015a) provide an excellent overview of recent pricing models under competition.

In the article by Gallego, Wang (2014) the authors consider a continuous time multi-product oligopoly for differentiated perishable goods. They use optimality conditions to reduce the multi-dimensional dynamic pure pricing problem to a one-dimensional one. Gallego, Hu (2014) analyze structural properties of equilibrium strategies in more general oligopoly models for the sale of perishable products. The solution of their model is based on a deterministic version of the model. Martinez-de-Albeniz, Talluri (2011) consider duopoly and oligopoly pricing models for identical products. They use a general stochastic counting process to model customer's demand. Further related models are studied by Yang, Xia (2013), Wu, Wu (2015), and Schlosser (2017). Dynamic pricing models under competition that also include strategic customers are analyzed by Levin et al. (2009) and Liu, Zhang (2013).

In most existing models, the demand intensity is assumed to be known. Dynamic pricing competition models with \textit{limited} demand information are analyzed by Tsai, Hung (2009), Adida, Perakis (2010), Chung et al. (2012), and Den Boer (2015b) using robust optimization and learning approaches. 
For a more comprehensive review, we refer to Chen, Chen (2015).

In contrast to the assumptions of most publications, in real-life applications, specific information is not observable, demand and price reactions 
are typically unknown, and customers as well as sellers might not act rational. Moreover, when dealing with dynamic pricing competition models, the most critical problem is their high complexity and the enormous number of potential market situations (cf. curse of dimensionality). 
Thus, most solution approaches are only applicable if specific assumptions can be verified and the number of competitors is small.

In general, dynamic pricing competition problems with incomplete information are \textit{not tractable} and cannot be optimally solved. Hence, in real-life applications, heuristic strategies have to be used. 
In practice, rule-based strategies as well as data-driven strategies are used, cf. Popescu (2016). 
As the evaluation of complex competing pricing strategies is also not analytically tractable, it is impossible to give performance guarantees of a specific strategy -- even if the underlying customer behavior as well as the competing strategies are known. 
Hence, simulation studies have to be used to compare the performance of dynamic pricing strategies and to analyze their strategic interaction in different oligopoly settings, see, e.g., Kephart et al. (2000) or Serth et al. (2017). 

\subsection{Contribution}

We tackle the problem of identifying efficient pricing strategies that are applicable in real-life scenarios.
%
In this paper, we present the following contributions: 

(i) \hspace{0.15cm}
We show how to compute dynamic price adjustments to be applied in practice (Sec. 3).

(ii) \hspace{0.05cm}We demonstrate that our heuristic is applicable even if the number of competitors is large and the 

\hspace{0.7cm}competitors' strategies are unknown (Sec. 4).

(iii) We verify the performance of our pricing strategy by comparing it to upper bounds, which are

\hspace{0.7cm}obtained by optimal strategies that take advantage of price anticipations (Sec. 5).

(iv) We find that higher price adjustment frequencies can easily overcompensate a loss in expected profits

\hspace{0.7cm}due to the lack of price anticipations.

(v)\hspace{0.11cm} We successfully applied our strategy on the Amazon Marketplace (Sec. 6).

\medskip

We present a dynamic programming model that allows for high dimensional market situations characterized by offers of several competitors with multiple features.
We show how real-life market data consisting of competitors' offers, our offers, and our sales can be used to estimate sales probabilities for arbitrary 
market situations.
This way, the impact of market dynamics and competitors' price reactions are implicitly taken into account 
without anticipating future market situations explicitly. 
Further, we use a decomposition approach to compute pricing decisions for single situations separately.
This makes it possible to efficiently compute price adjustments that are based on
current market situations. 
%
%
%
Finally, we compensate the lack of perfect price anticipations by frequent price adjustments, which in turn are possible as the model allows for rapid re-computations of optimized prices.
Further, we present an extended model which takes price adjustment costs into account. This allows to combine fast reaction times and a limited use of price updates.

Our approach is successfully applied in practice. The model was calibrated using 
historical market data from the Amazon Marketplace. Results show that our strategy outperforms
the established rule-based strategy of an experienced seller regarding profitability \textit{and} speed of sales. 


\section{Model Description}

We consider the situation where a firm wants to sell a finite number of perishable goods on a digital market platform (e.g., Amazon or eBay). For the products, there are several competitors. In our model, we allow that customers compare prices. They also might take additional product features, such as qualities or ratings of different competitors into account.

We assume that items cannot be reproduced or reordered. The initial number of items to sell is $N$, $N \le \infty$, and the time horizon is finite. If a sale takes place, shipping costs $c$ have to be paid, $c \ge 0$. A sale of one item at price $a$ leads to a net revenue of $a-c$. Moreover, we consider inventory holding costs. We assume that each unsold item leads to inventory costs of $l$ per period (e.g., one hour or one day), $l \ge 0$. Furthermore, we use the discount factor $\delta$, $0 < \delta  \le 1$, for a period of length one.

In our discrete time model, we consider sales probabilities 
denoted by $P$. Due to customer choice, sales probabilities will particularly depend on our offer price $a$ and the competitors' prices. Moreover, we allow sales probabilities to depend on time. In our model, $P$ is a general function of our offer price $a$ and a market situation denoted by $\vec s$. The market situation $\vec s$ is a vector which includes all relevant observable quantities of interest, such as time $t$, the competitors' prices $\vec p$, customer ratings $\vec v$, product conditions, etc. (see, e.g., Kachani, Shmatov (2010)). The probability to sell exactly $i$ items within the time span $(t,t + 1)$ in a \textit{stable} market situation $\vec s = (t,\vec p,\vec v,...)$ is denoted by, $t = 0,1,2,...,T-1$, $a \ge 0$, $i=0,1,2,...$,
\begin{equation}
\label{2.1}
{P_t}(i,a,\vec s).
\end{equation}

In real-life scenarios the sales probabilities ${P_t}(i,a,\vec s)$ are typically not known and have to be estimated using data-driven approaches. 
In simulation models, the sales probabilities ${P_t}(i,a,\vec s)$ can be specifically defined 
and allow to verify the quality of data-driven estimations as well as the performance of pricing strategies.
For instance, assuming a given sales intensity $\lambda  = {\lambda _t}(a,\vec s)$ and a scaling factor $d$, we can define Poisson distributed sales probabilities for various intervals, $t = 0,1,2,...,T-1$, $i = 0,1,2,...$, $a \ge 0$, $d>0$,

\[\blau{{P_t}(i,a,\vec s): = Pois\left( {d \cdot {\lambda _t}(a,\vec s)} \right).
}\]

\smallskip
The random inventory level at time $t$ -- the beginning of period $t$ -- is denoted by ${X_t}$, $t = 0,1,...,T$.
The end of sale (for our firm) is either time $T$ or the random time $\tau$, when all $N$ products are sold.
%
%
For each period $t$, a new offer price $a$ can be chosen. We call pricing strategies admissible if they are non-anticipating (Markov policies); pricing decisions ${a_t} \ge 0$ may depend on time $t$, the own inventory level ${X_t}$, and the current market situation (denoted by ${\vec S_t}$), which particularly contains the current prices of the competitors. A list of variables and parameters is given in the Appendix, cf. Table A.7.

The firm's profits are defined by each period's realized sales (cf. $X_{t} - X_{t+1}$) and inventory holding costs (cf. $l \cdot X_{t}$). For chosen prices $a_t$, the random accumulated profit $G_t$ from time $t$ on (discounted on time $t$) amounts to, $t = 0,1,...,T$,
\begin{equation}
\label{2.3}
{G_t}: = \sum\limits_{s = t}^{T-1} {{\delta ^{s - t}} \cdot \left( {({a_s} - c) \cdot \left( {{X_{s}} - {X_{s+1}}} \right) - l \cdot {X_{s}}} \right)}.
\medskip
\end{equation}
We look for a pricing policy 
that maximizes expected (discounted) future profits $E({G_t}\left| {{X_t} = n,} \right.{\vec S_t} = {\vec s_t})$ for all times $t$, inventories $n$, and occurring market situations ${\vec s_t}$.

In the following sections, we will solve dynamic pricing problems that are related to \eqref{2.1} - \eqref{2.3}. In the next section, we propose our heuristic pricing strategy for scenarios with many competitors and unknown strategies. The applicability of our strategy is demonstrated in Section 4. In Section 5, we measure the performance of our pricing strategy by studying various settings in which competitors frequently and strategically adjust their prices.
In Section 6, our approach is used in a live production setting on Amazon.



\section{Circumventing the Curse of Dimensionality: A Heuristic Approach}

In this section, we derive viable pricing strategies for markets with many competitors.
We use an efficient algorithm to circumvent the curse of dimensionality and propose an effective heuristic pricing strategy.
%
%

There are two major problems to derive applicable dynamic programming strategies in competitive markets: 
(i) as demand is affected by many parameters (e.g., dozens of competitors' prices) a model's state space explodes, and 
(ii) in general, competitors' strategies are not known, and competitors' price adjustments and other market evolutions
cannot be fully anticipated.

Our approach deals with both problems. 
Most importantly, we subsequently compute prices for \textit{one period} only based on the current market situations that occur during a sales process.
To compute \blue{a price for the next time period}, in general, the current market situation (current state) as well as potential evolutions of the market (future states) have to be taken into account.
As price reactions of competitors occur with a certain delay, the \textit{short-term} evolution of the market can be well approximated by the current market situation. The \textit{long-term} evolution of the market, however, can hardly be predicted. 

In our use-case, the optimal price for one period mostly depends on the current state and is less affected by possible market evolutions in the far future. Note, while the expected profit of the next period highly depends on the current price decision and the current market situation, the expected future profits are typically much less influenced by the current price decision. This is due to the following: First, future profits are determined by various upcoming decisions that account for future states. Second, our price update typically hardly affects the long-term evolution of the market (especially if many firms are involved). Third, in markets in which price patterns are cyclic or fluctuate around certain reference prices, it can be assumed that the best expected long-term profits are similar for different market situations. 

Hence, instead of trying to predict possible market evolutions in the future, in our approach, we compute price updates based on an accurate calculation of short-term profits and expected long-term profits that are approximated based on current market conditions.

For a current state, we manage problem (i) as follows:
We roughly approximate future market situations by using sticky prices. 
While the degree of inaccuracy is acceptable, we gain a \textit{structure} that
makes it possible to circumvent the curse of dimensionality, cf. problem (i), 
as the states of our dynamic system (i.e., the market situation) are not coupled and can be easily decomposed.
Thus, decisions for single market situation can be computed independently from others.
 
The second key idea is to compensate the ``sticky'' model's inaccuracy as well as the lack of correct price anticipations, cf. problem (i), by \textit{frequent} price adjustments, which in turn, are possible as the model's simplicity allows for fast re-computations.


Due to price adjustments, the exit, or entry of firms, in general, market situations are not stable. In our model, we differentiate between sales probabilities ${P_t}(i,a,\vec s)$ for \textit{stable} market situations, cf. \eqref{2.1}, and \textit{conditional} sales probabilities, $t = 0,1,...,T-1$,  $a \ge 0$, $i = 0,1,2,...$,
\begin{equation}
\label{2.2}
{\tilde P_t}(i,a|\vec s)
\end{equation}
for selling $i$ items within \blue{the time interval} $(t,t+1)$ at price $a$ under the condition that at time $t$ the market situation is $\vec s$ \blue{(but \textit{may change} within the period)}. Note, the probabilities ${\tilde P_t}(i,a|\vec s)$, cf. \eqref{2.2}, can be estimated from real-life data \blue{where market situations are typically not stable}.

As our approach to compute price adjustments is designed to be applied in unstable markets, it will be based on (estimated) conditional probabilities ${\tilde P_t}(i,a|\vec s)$\blue{, cf. (3)}. As described in the beginning of the section, we use a simplified dynamic programming approach in which expected future profits $E({G_t}|{X_t} = n{,_{}}{\vec S_t} = \vec s)$, cf. \eqref{2.3}, are described by the value function ${V_t}(n,\vec s)$, $t = 0,1,...,T$, $n = 0,1,...,N$.
If all items are sold or the time horizon is over, no future profits can be made, i.e., for any \blue{market situation} $\vec s$ the natural boundary conditions of the value function are given by, $n = 0,1,...,N$, $t = 0,1,...,T$,
\begin{equation}
\label{4.1}
V_T^{}(n,\vec s) = 0,
\end{equation}
\begin{equation}
\label{4.2}
V_t^{}(0,\vec s) = 0.
\end{equation}

The remaining values ${V_t}(n,\vec s)$ \blue{are given by 
the Hamilton-Jacobi-Bellman (HJB) equation}, $n = 1,...,N$, $t = 0,1,...,T-1$,
\begin{equation}
\label{4.3}
{V_t}(n,\vec s) = \mathop {\max }\limits_{a \in A} \left\{ {\sum\limits_{i \ge 0} {{{\tilde P}_t}(i,a|\vec s)}  \cdot \left( {(a - c) \cdot \min (n,i) - n \cdot l + \delta  \cdot {V_{t + 1}}\left( {{{(n - i)}^ + },\vec s} \right)} \right)} \right\}
\end{equation}
and can be computed recursively. The set of admissible prices $A$ can be continuous or discrete.
Finally, the pricing strategy is determined by the arg max of \eqref{4.3}, $n = 1,...,N$, $t = 0,1,...,T-1$,
\begin{equation}
\label{4.4}
{a_t}(n,\vec s) = \mathop {\arg \max }\limits_{a \in A} \left\{ {\sum\limits_{i \ge 0} {{{\tilde P}_t}(i,a|\vec s)}  \cdot \left( {(a - c) \cdot \min (n,i) - n \cdot l + \delta  \cdot V_{t + 1}^{}\left( {{{(n - i)}^ + },\vec s} \right)} \right)} \right\}.
\end{equation}

\noindent In case the prices determined by \eqref{4.4} are not unique, we choose the largest one. 

Note, due to the size of the state space it is not possible to compute prices ${a_t}(n,\vec s)$ for \textit{all} states $\vec s$ in advance. 
The following algorithm, however, circumvents the curse of dimensionality \blue{by considering just single states and computing feedback strategies multiple times. This allows to derive} viable heuristic pricing strategies in competitive markets with a large number of competitors. 

\bigskip

\noindent \textbf{\blue{Algorithm 3.1}}

We propose the following pricing heuristic:
\smallskip

(Step 1)	\hspace{0.19cm} For the current point in time $t$, observe the new state, i.e., the current inventory level ${X_t}$

\hspace{1.7cm}and the current market situation ${\vec S_t}$.

(Step 2)	\hspace{0.18cm} Use $T - t$ recursion steps and the probabilities ${\tilde P_s}(i,a|{\vec S_t})$, $s=t,...,T-1$, $i = 0,1,...,{X_t}$, 

\hspace{1.72cm}$a \in A$, to compute the specific value ${V_t}({X_t},{\vec S_t})$, cf. \eqref{4.3}.

(Step 3)	\hspace{0.22cm} Choose price $a_t^{}({X_t},{\vec S_t})$, which is associated to the last step of the recursion, cf. \eqref{4.4}.

\hspace{1.72cm}At the point in time of the next price adjustment, i.e., $t+1$, go to Step (1) and re-solve the 

\hspace{1.72cm}system \blue{(6) - (7) based on the new inventory $X_{t+1}$ and} the new state $\vec S_{t+1}$. 

\medskip

\blue{We just have to} compute prices for single (current) market situations and to regularly refresh prices in response to changing market situations. Note, a single recomputation \blue{of (6) - (7)} is very fast and does neither increase with the number of competitors nor increase with the number of dimensions of the market situation.


\section{Application of the Heuristic Strategy}

\subsection{Sales Probabilities}

In this section, we assume that market situations are \blue{only} characterized by the point in time $t$ and the prices of $K$ competitors. The vector of prices $\vec p := (p_1,...,p_K)$ does not necessarily have to be ordered.
To be able to demonstrate the applicability and the performance of pricing strategies, cf. next Section 5, we need to quantify realistic sales probabilities in competitive markets. Instead of using made-up or highly stylized demand probabilities (e.g., of linear, exponential, or isoelastic type), we seek to use a more general (data-driven) demand setting. As an example, we define such probabilities based on a logistic regression model using linear combinations of state-dependent features/regressors $\vec x = \vec x(a,\vec s)$ and given coefficients $\vec \beta $; i.e., we consider binary probabilities of the form, $t = 0,1,...,T$, $a \ge 0$, $i \in \{0,1\}$,
\[{\tilde P_t}(i,a|\vec s):= i \cdot \hat P(a,\vec s) + (1-i) \cdot  (1-\hat P(a,\vec s)),\]
where the logit probability $\hat P(a,\vec s)$ is given by
\begin{equation}
\label{4.5}
\hat P(a,\vec s) := {\frac{{{e^{\vec x(a,\vec s)'\vec \beta }}}}{{1 + {e^{\vec x(a,\vec s)'\vec \beta }}}}}.
\end{equation}


\blue{The dependent variable is the number of sales within a certain time frame}. In the following definition, we give simple examples of explanatory variables. \blue{In this framework}, further explanatory variables can be easily defined to capture the impact of various effects, such as customer ratings, product quality, shipping time, the type of product (categories/clusters), etc.

\smallskip

\begin{definition}
\blue{Let market situations be described by $\vec s = (t,\vec p)$. For potential offer prices $a$}, we define the following regressors $\vec x = \vec x(a,\vec s)$:

(i)\hspace{0.4cm}	 $x_1(a,\vec s) := 1$ \hspace{3.2cm}	\blue{constant/intercept}

(ii)\hspace{0.3cm}	 $x_2(a,\vec s):= r(a,\vec p)$	\hspace{2.4cm} rank of price $a_{}$ within the prices $\vec p_{}$, where
\[ \blue{ r(a,\vec p):= 1 + card ({\left\{ {k = 1,...,K_{}\left| {p_k < a} \right.} \right\}}) + 0.5 \cdot card ({\left\{ {k = 1,...,K\left| {p_k = a} \right.} \right\}}) } \]

(iii)\hspace{0.2cm}	 $x_3(a,\vec s):= a_{} - \mathop {\min }\limits_{k = 1,...,K_{}} \{ p_k\} $ \hspace{0.87cm}	price gap between \blue{price} $a_{}$ and the best competitor's price

(iv)\hspace{0.23cm}	 $x_4(a,\vec s):= K_{}$ \hspace{3.1cm}	total number of competitors for product $i$ in period $t$

(v)\hspace{0.32cm}	 $x_5(a,\vec s):= (a_{} + \sum\nolimits_k {p_k} )/(1 + {K})$	 \hspace{0.2cm} average price level for product $i$ in period $t$

(vi)\hspace{0.25cm}	 $x_6^{}(a,\vec s): = {\alpha _1} \cdot {t^{{\alpha _2}}}$	 \hspace{2.3cm} time dependence, e.g., via scaling parameters $\alpha_1$, $\alpha_2$
\label{def1}
\end{definition}


\begin{example}
\blue{Let market situations} be characterized by $\vec s = \vec p$. We consider the five variables $x_m(a,\vec p)$, $m = 1,...,5$, see Definition \ref{def1} (i)-(v). Inspired by a production data set obtained from Amazon, see Section 6, we let
\[\vec \beta  = (\beta _1^{},\beta _2^{},\beta _3^{},\beta _4^{},\beta _5^{}) = ({\rm{ - 3}}{\rm{.89}}{,_{}}{\rm{ - 0}}{\rm{.56}}{{\rm{,}}_{}}{\rm{ - 0}}{\rm{.01}}{{\rm{,}}_{}}{\rm{0}}{\rm{.07}}{{\rm{,}}_{}}{\rm{ - 0}}{\rm{.05}}).\]
\blue{In this context}, we define binary sales probabilities for a given market situation $\vec p$ \blue{and offer prices $a$} via
%
$\hat P(a,\vec p):= {e^{\vec x(a,\vec p)'\vec \beta }}/(1 + {e^{\vec x(a,\vec p)'\vec \beta }})$, cf. \eqref{4.5}.
In this example, we consider the following market situation \blue{with $K=10$ competitor prices}:

\[\vec p = \left( {{{5.18, }_{}}{{5.96, }_{}}{{6.31, }_{}}{{8.28, }_{}}{{9.48, }_{}}{{9.88, }_{}}{{10.33, }_{}}{{10.98, }_{}}{{11.67, }_{}}13.52} \right).\]
\label{ex1}
\end{example}

\vspace{-0.5cm}


The demand coefficients defined in Example \ref{ex1} are based on a large data set ($>$20 M observations/month) from the Amazon market for used books including $100\,000$ different products and up to $K=20$ competitors, 
cf. Schlosser et al. (2016). The average length of a period is 2.4 hours. 

Given the market situation $\vec p$ described in Example \ref{ex1}, Figure 1a illustrates sales probabilities $\hat P(a,\vec p)$ for different potential offer prices $a$. \blue{As expected, we observe that the sales probability decreases with the price $a$, cf. Definition 4.1 (ii) and $\beta_2:=-0.56<0$}. The jumps in Figure 1a occur whenever the price rank changes. Note, undercutting a \blue{competitor's price} yields a better price rank, which can easily double the sales probability.
The corresponding expected profits $(a - c) \cdot \hat P(a,\vec p)$ for one period of time are shown in Figure 1b. Expected profits are positive as long as $a$ exceeds the shipping costs $c$. The peaks in Figure 1b are caused by the price rank related jumps of $\hat P(a,\vec p)$. In this example, expected profits are maximized if one slightly undercuts the \blue{best competitors' price, i.e., $p_1 = 5.18$}.


\begin{figure}[ht]
\begin{center}
\includegraphics[scale=0.5]{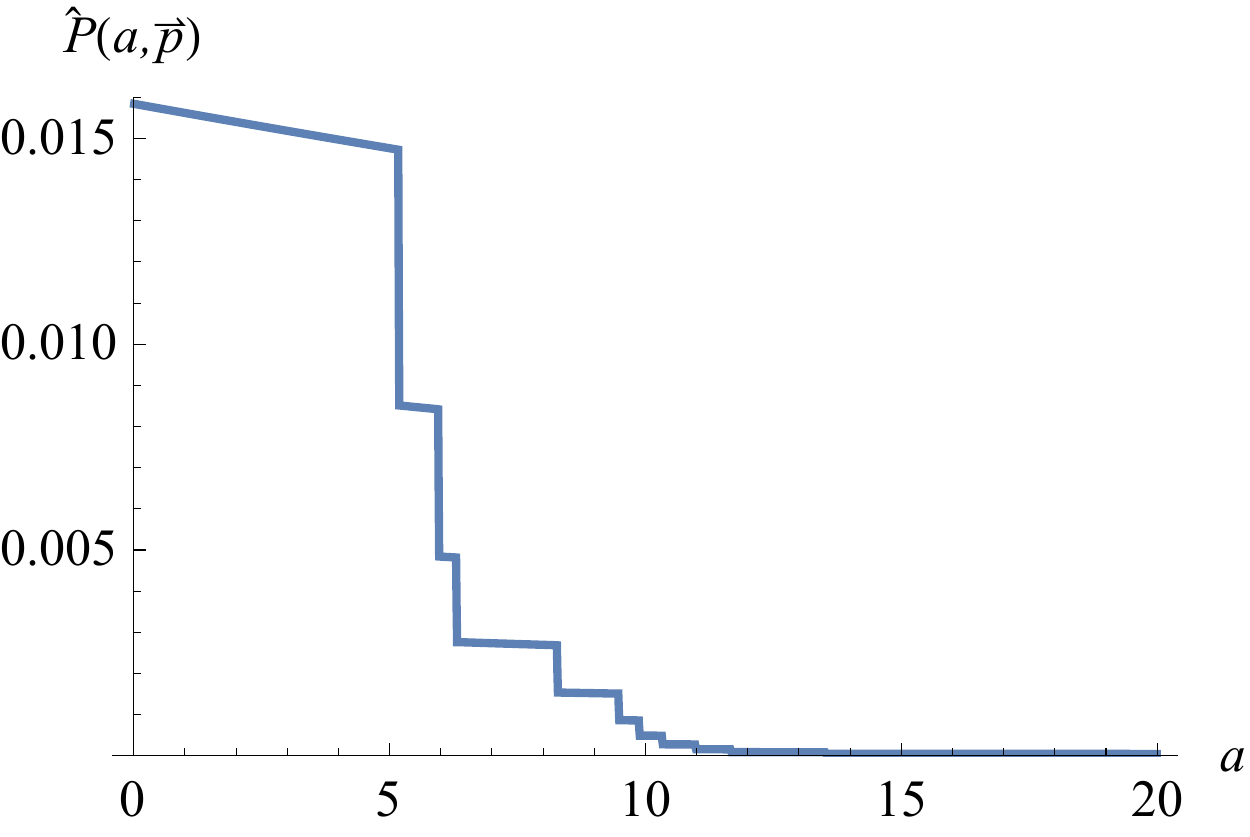}\qquad
\includegraphics[scale=0.5]{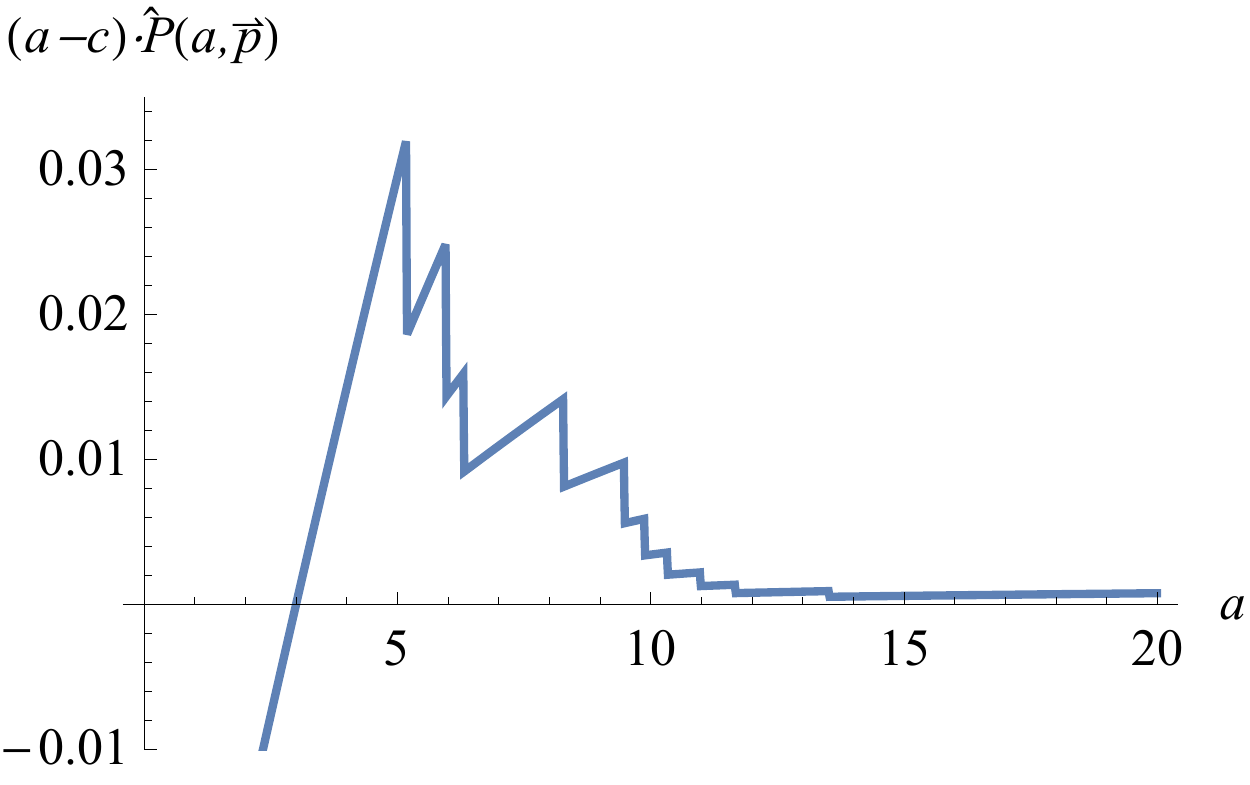}
\caption{Predicted sales probabilities  (left window 1a) and expected profits for one period (right window 1b) \blue{for different potential offer prices $a \in A:= \left\{ {0.01,0.02,...,20} \right\}$, shipping costs $c = 3$}; Example \ref{ex1}.}
\label{fig1}
\end{center}
\end{figure}



\begin{remark}   (Model calibration and data-driven demand estimations) 

Our approach to model sales probabilities is simple yet reasonable, especially for practitioners. Logistic regression is a well-understood method and allows to handle hundreds of features and millions of observations. Regression results are robust and can be directly interpreted. Nevertheless, our model also allows to calibrate sales probabilities using other approaches (e.g., using decision trees, see Quinlan (1986), or gradient-boosted decision trees, see Chen, Guestrin (2016), etc.).
\label{rem3}
\end{remark}

\subsection{Numerical Examples}

In the following, we demonstrate the applicability of our repricing approach in competitive markets, cf. Algorithm 3.1. We illustrate the computation of the heuristic strategy using the sales probabilities defined in Section 4.1.
\smallskip

\begin{example}
We consider the setting of Example \ref{ex1}. We let $T=100$, $c=3$, $\delta=0.9995$, $N=25$, $l=0.01$, and assume the exemplary state $\vec s = \vec p = \left( {{{5.18,}_{}}{{5.96,}_{}}{{6.31,}_{}}{{8.28,}_{}}{{9.48,}_{}}{{9.88,}_{}}{{10.33,}_{}}{{10.98,}_{}}{{11.67,}_{}}13.52} \right)$. We illustrate state-dependent prices \blue{of our heuristic}, cf. \eqref{4.3}-\eqref{4.4}, in case of the Poisson probabilities ${\tilde P_t}(i,a|\vec s):= Pois( {d \cdot \hat P(a,\vec p)} )$, $i=0,1,...$, $d=10$, $a \in A:= \left\{ {0.01,0.02,...,20} \right\}$.
\label{ex2}
\end{example}

\vspace{0.3cm}

\begin{figure}[ht]
\begin{center}
\includegraphics[scale=0.45, trim = 22mm 0mm 0mm 0mm, clip]{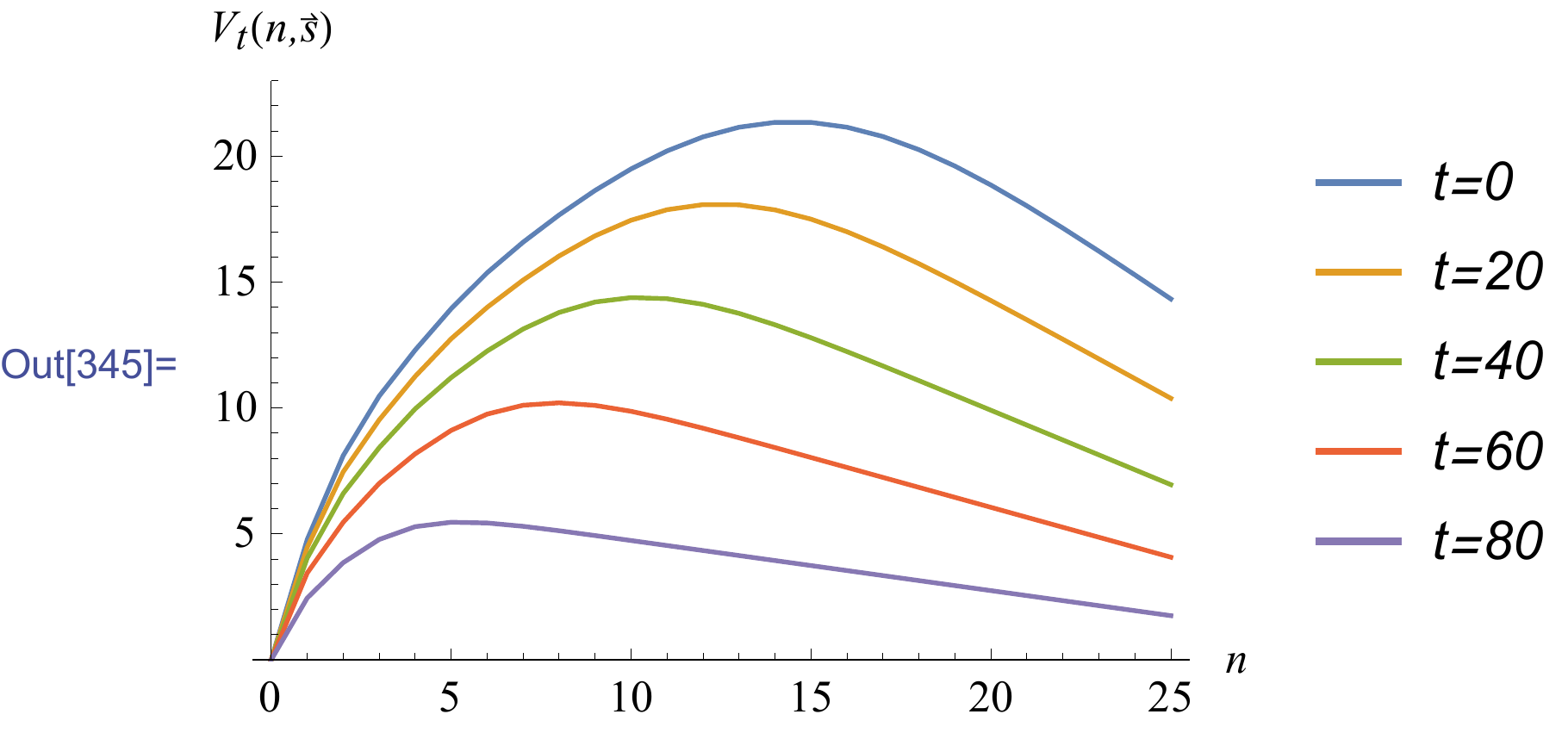}\qquad
\includegraphics[scale=0.45, trim = 22mm 0mm 0mm 0mm, clip]{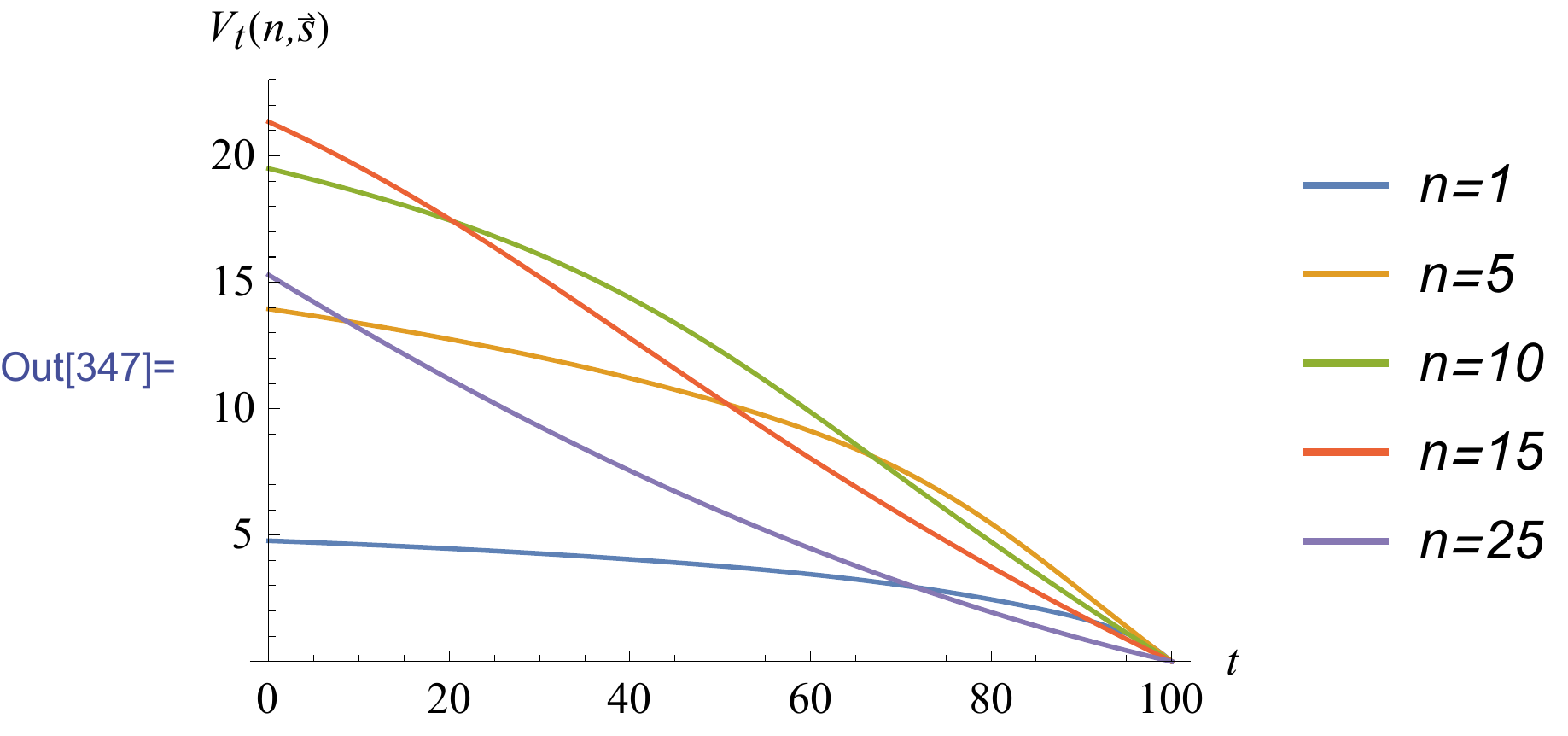}
\caption{\blue{Illustration of the value function for different points in time $t$ (left window 2a) and for different inventory levels $n$ (right window 2b) given a specific market situation, $t=0,...,100$, $n=1,...,25$}; Example \ref{ex2}.}
\label{fig2}
\end{center}
\end{figure}
\vspace{-0.1cm}

The value function (Figure 2) and the associated heuristic feedback prices (Figure 3) are illustrated for the setting of Example \ref{ex2}. 
\blue{Figure 2a and 2b summarize (approximated) expected profits at a \textit{current} time $t$ in state $n$ given the market situation $\vec s$}. As inventory holding costs are included in the model it is not advantageous to store too many items.
\blue{In this example, we observe that at time $0$ one should not store more than 15 items.
If 20 periods of time are left ($t=80$) the best expected profit can be gained by offering 5 items. 
Having more items in stock is unfavourable due to inventory holding costs.
}

Figure 3 illustrates prices that are recommended by our heuristic at a \textit{current} time $t$ in state $n$ given the market situation $\vec s = \vec p$.
We observe that prices are higher the less items are left to sell and the more time is left. For instance,
if more than 50 periods of time are left and there is only one item left to sell the price is \blue{$9.47 = p_5 - 0.01$ (rank 5)}.
If more than 60 periods of time are left and there are 2 or 3 items left to sell the price is 8.27 (rank 4).
If more than 3 items are left to sell the price is either 5.95 (rank 2) or 5.17 (rank 1) depending on the time-to-go.
If more than 7 items are left in stock the price ${a_t}(n,\vec s)$ will occupy price rank 1 at every point in time. 
Note, price rank 3 (6.30) is not chosen at all since the difference between price rank 2 and 3 is too small.

\begin{figure}[ht]
\begin{center}
\includegraphics[scale=0.4, trim = 22mm 0mm 0mm 0mm, clip]{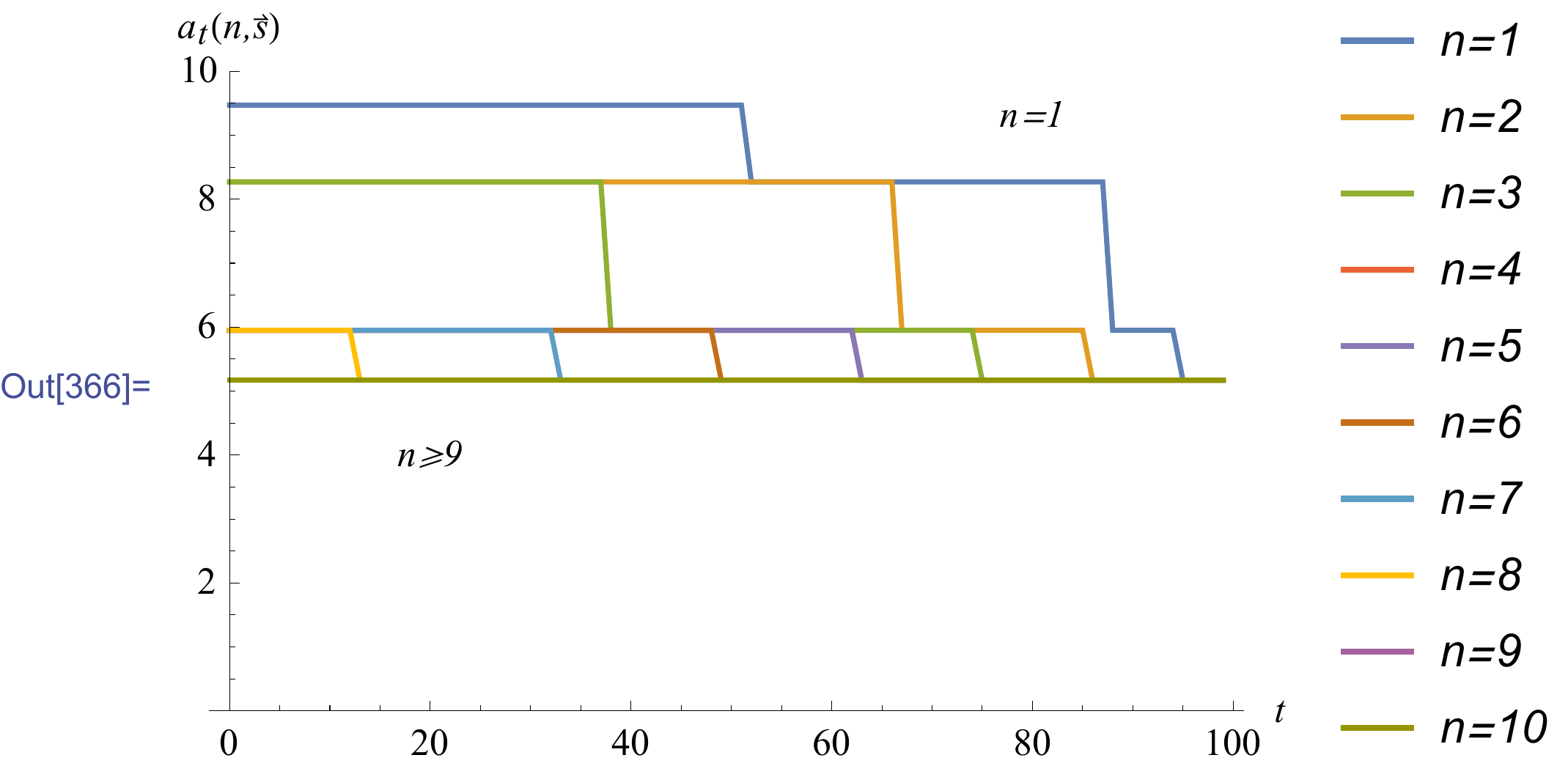}
\caption{\blue{Heuristic feedback pricing policy for a specific market situation, $t=0,...,100$, $n=1,...,10$}; Example \ref{ex2}.}
\label{fig3}
\end{center}
\end{figure}
\vspace{-0.15cm}

Next, we demonstrate the \textit{applicability} of our approach in competitive markets with many active competitors and unknown strategies. 
We consider a numerical example that can be reproduced by the reader.

\smallskip

\begin{example}
We consider the setting of Example \ref{ex2}, i.e., we let $T=100$, $c=3$, $\delta=0.9995$, $N=10$, $d=10$, $l=0.01$. 
We simulate competitors' price trajectories over time as follows. 
We let competitors adjust their prices using random reaction times and randomized price changes. The probability that a single firm $k$ adjusts its price $p_t^{(k)}$, $k=1,...,K$, at $t = 0,h,...,T - h$, $h$=0.1, is denoted by $\pi$, $\pi  = 0.3 \cdot h$. 
\blue{The random amplitude of price jumps are normalized via $h$, $T$, and $\pi$ such that average trends over time are the same even if the frequency of price jumps (cf. $\pi$) or the period length $h$ are different.}
Initial prices ${\vec p_0}$ are chosen as in Example \ref{ex2}. We simulate price trajectories using four scenarios of randomized price adjustments, cf. case (i) -- (iv), $K=10$, $t = 0,h,...,T - h$:
\label{ex3}

\medskip

(i)\hspace{0.38cm} 	no price trend: \hspace{0.95cm} $p_{t + h}^{(k)} = \max \left( {c + 0.01,p_t^{(k)} + {1_{\{ U(0,1) < \pi \} }} \cdot U( - 20,20) \cdot h/\pi /T} \right)$

(ii)\hspace{0.28cm} 	positive price trend: \hspace{0.18cm} $p_{t + h}^{(k)} = \max \left( {c + 0.01,p_t^{(k)} + {1_{\{ U(0,1) < \pi \} }} \cdot U( - 15,25) \cdot h/\pi /T} \right)$

(iii)\hspace{0.16cm} 	negative price trend:  \hspace{0.11cm} $p_{t + h}^{(k)} = \max \left( {c + 0.01,p_t^{(k)} + {1_{\{ U(0,1) < \pi \} }} \cdot U( - 25,15) \cdot h/\pi /T} \right)$
\smallskip

(iv)\hspace{0.23cm} 	strategic competitor:\hspace{0.22cm}Consider setting (i), in which the other competitors' price adjustments are 

\hspace{0.9cm}mutual independent and have no trend. One single firm, however, plays a strategic price adjustment 

\hspace{0.9cm}strategy by steadily undercutting our price by $\varepsilon$=0.2.

\medskip

In this example, we also consider stock-outs of firms by simulating competitors to exit the market at random times. Similarly, we allow new competitors to enter the market with a random starting price. 
Our realized sales are simulated using sales probabilities for time intervals $(t,t+h)$, $t = 0,h,...,T - h$ of size $h$ defined by 
$P_t^{(h)}(i,a,\vec p):= Pois( {h \cdot d \cdot \hat P(a,\vec p)})$, cf. \eqref{4.5}.
Our firm adjusts its price at $t = 0,1,...,T - 1$ according to Algorithm 3.1; for ease of simplicity, we use ${\tilde P_t}(i,a|\vec s):= Pois( {d \cdot \hat P(a,\vec p)})$.

\end{example}

\smallskip


\begin{figure}[ht]
\begin{center}
\includegraphics[scale=0.32]{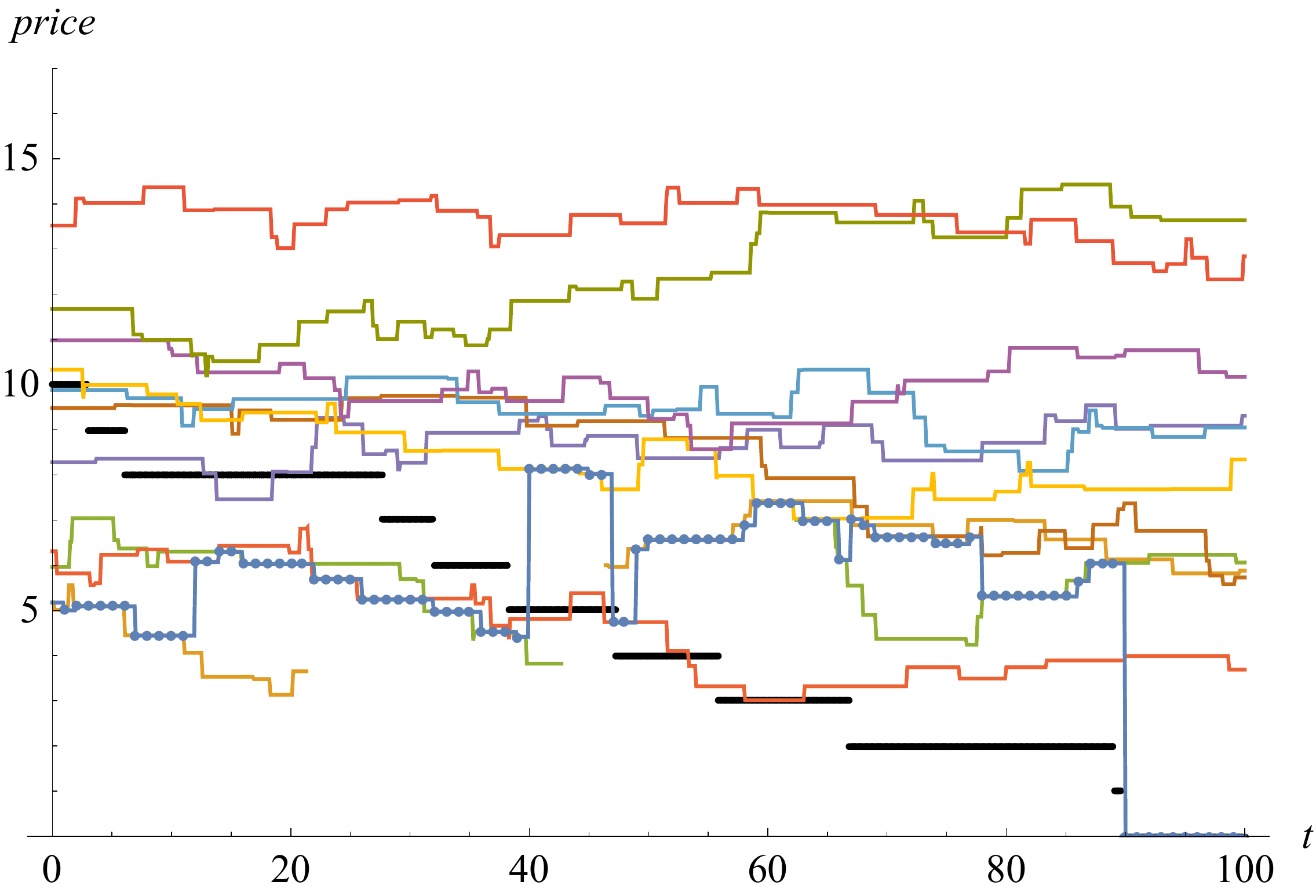}\qquad
\includegraphics[scale=0.32]{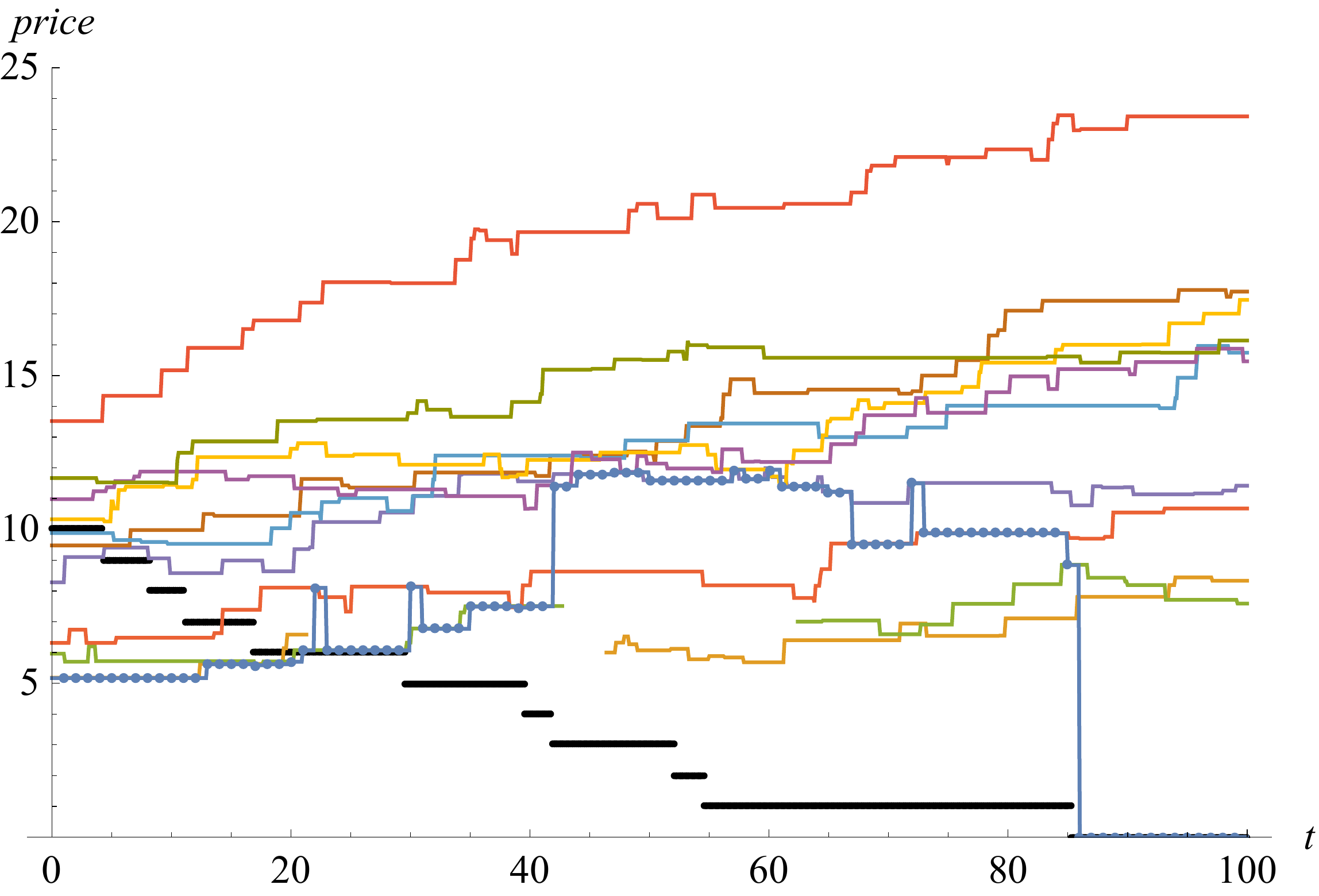}
\smallskip

\includegraphics[scale=0.32]{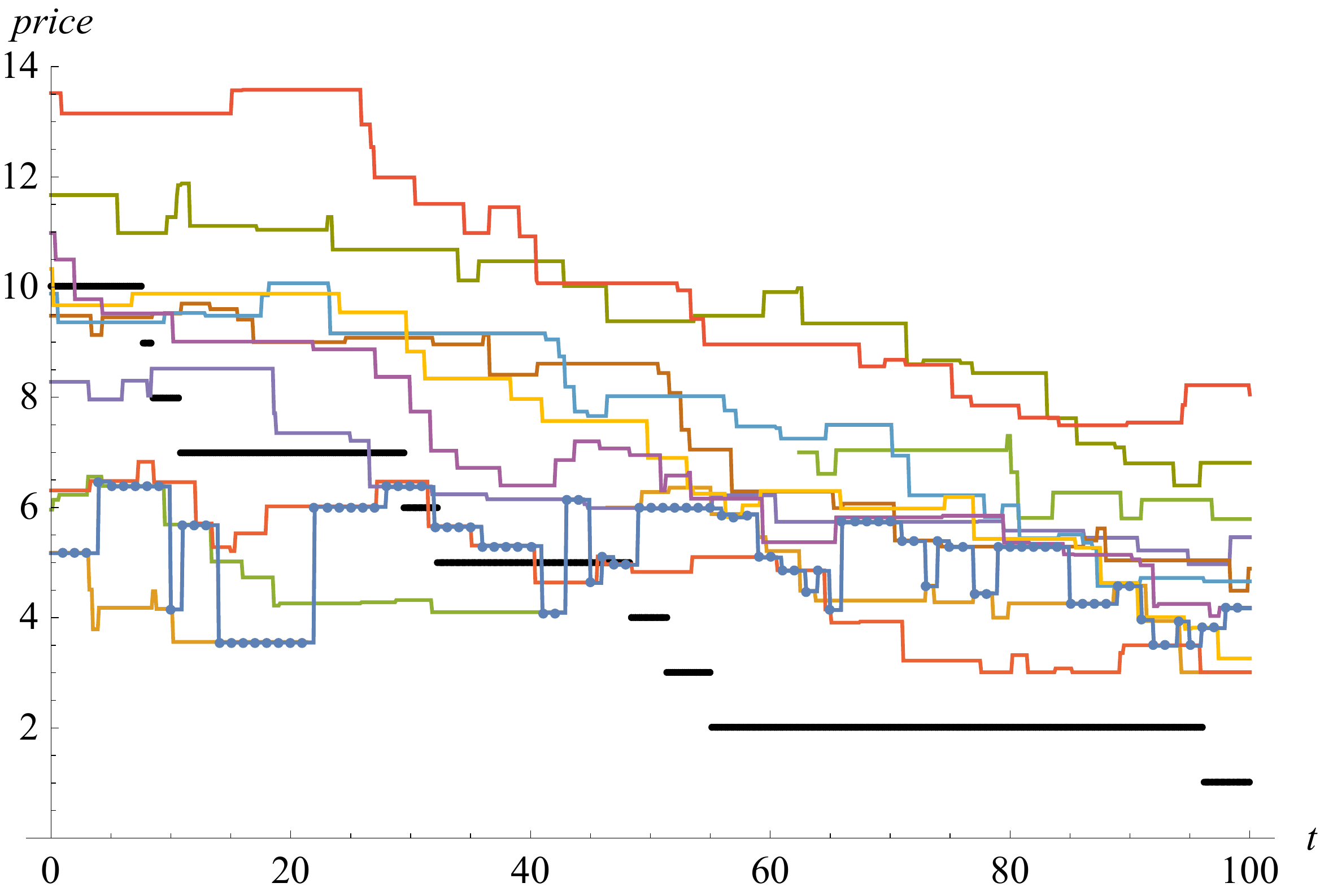}\qquad
\includegraphics[scale=0.32]{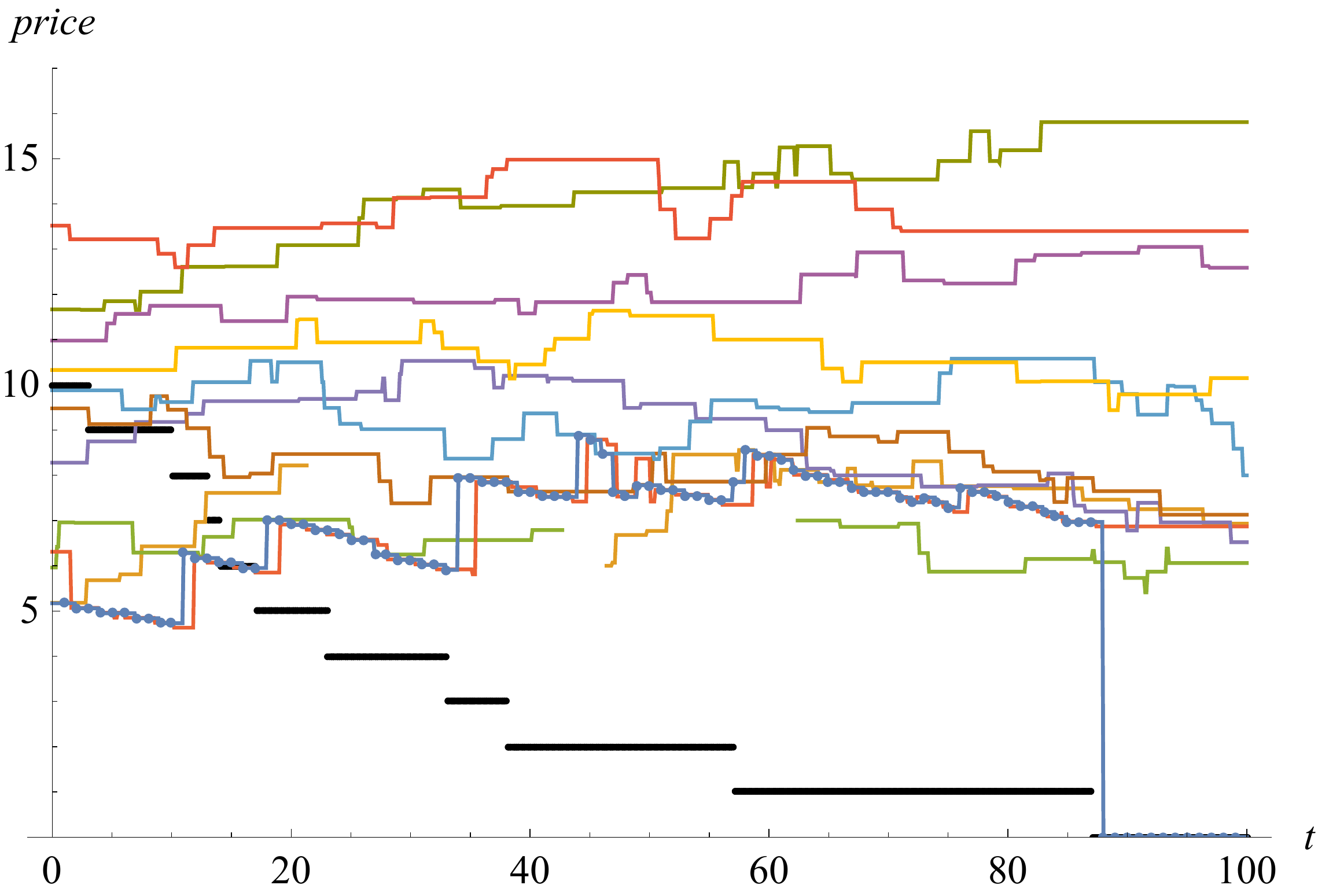}
\caption{Application of our heuristic dynamic pricing strategy under competition; scenario (i) (upper left window 4a), scenario (ii) (upper right window 4b), scenario (iii) (lower left window 4c) and scenario (iv) (lower right window 4d); Example \ref{ex3}.}
\label{fig4}
\end{center}
\end{figure}

For case (i) - (iv) of Example 4.3, the four subfigures of Figure 4 show price trajectories of the competing firms over time. The price trajectory of our firm can be identified by little blue bullet points indicating the price adjustments at times $t$, $t=0,1,...,T-1$, according to the strategy defined in Algorithm 3.1 for the observed states $\vec S_t$ and remaining inventory levels $X_t$. In this context, the black decreasing step function illustrates the remaining inventory level of our firm. 
The evaluation of each scenario takes less than a second.




In scenario (i)-(iii), we observe that the heuristic strategy slightly undercuts other competitors, cf., Figure 4a, 4b, and 4c. Furthermore, we observe two effects. First, if the prices of the competitors are high, then the strategy occupies leading price ranks, see Figure 4b; if competitors' prices are low, it is more profitable to use moderate price ranks, see Figure 4c. Second, at the end of the time horizon the strategy occupies leading price ranks.

Scenario (iv) illustrates the case of an aggressive strategic competitor. We observe that the interplay of the strategic competitor's and the price adjustments of our heuristic lead to specific cyclical price patterns, cf. Figure 4d, which have similarities to Edgeworth-price cycles, see, e.g., Maskin, Tirole (1988) or Noel (2007). Whenever the price level is too low, our firm raises the price up to a certain level. Note, the heuristic strategy just responds to market situations, prices are not anticipated.

Example \ref{ex3} demonstrates that our heuristic strategy is applicable for any scenario of the competitors' price trajectories -- even if the number of competitors is large and competitors enter or exit the market. The price adjustments of the heuristic depend on time, the inventory level as well as the current market situation.
Note, the number of computation steps of Algorithm 3.1 does not increase with the number of competitors. The approach remains also applicable if the competitors' price adjustments are mutually dependent. The repeating price adjustments of our firm allow to respond to any change of the market environment.

Moreover, the approach allows to take additional factors, such as ratings, product conditions, or shipping times into account. These extensions will only affect the sales probabilities of the model, e.g., by using additional explanatory variables, cf. Definition 4.1. The simple optimization of the model, however, is not affected, since the number of computation steps of Algorithm 3.1 does not increase. Our approach allows to directly include dozens of features and dozens of competitors without increasing the model's complexity.

The following remark summarizes important managerial recommendations.


\begin{remark}	(Managerial recommendations)  

(i)\hspace{0.35cm} If market conditions can be assumed to be overall stable, a reasonable initial inventory can be

\hspace{0.9cm}determined based on $\mathop {\max }\nolimits_{n \ge 0} \left\{ {{V_0}(n,\vec s)} \right\}$, cf. economic order quantities.

(ii)\hspace{0.27cm} It is mostly optimal to undercut one of the competitors' prices by the smallest unit (e.g., 1 cent). 

\hspace{0.9cm}In this context, Algorithm 3.1 can be accelerated dramatically by considering just a small subset of 

\hspace{0.9cm}admissible prices, i.e.,
\[
\tilde A(\vec p):= \bigcup\limits_{k = 1,...,K} {\left\{ {{{({p_k} - \varepsilon )}^ + }} \right\}}.
\]

(iii)\hspace{0.25cm}As expected, discounting motivates short-term profits and leads to more aggressive, i.e., lower 

\hspace{0.9cm}feedback prices. Hence, a lower (higher) discount factor $\delta$ can be used to increase (or decrease) 

\hspace{0.9cm}the overall aggressiveness of the strategy. This "instrument" can also be used to actively manage 

\hspace{0.9cm}the total inventory of a firm as well as its cash flow. If negative (or positive) price trends can be 

\hspace{0.9cm}anticipated, cf. Fig. 4b (and 4c), the aggressiveness of the strategy can also be used to further

\hspace{0.9cm}improve the heuristic strategy.
\label{rem4}
\end{remark}






\section{Measuring the Performance of the Heuristic Strategy}

In this section, we want to measure the \textit{performance} of our strategy derived in Section 3.
We consider the case of many active competitors, cf. Section 5.1, as well as the case of strategic competitors that strategically respond to our price adjustments, cf. Section 5.2.
In Section 5.3, we analyze the case in which our strategy \blue{is played against itself}.



\subsection{Heuristic vs. Optimal Forward-Looking Strategies in Dynamic Oligopoly Competition}

\blue{We measure the performance} of our heuristic by comparing its expected profits to \textit{upper bounds} determined by optimal expected profits for the case that the competitors' future prices are \textit{\blue{completely} known} in advance. In this subsection, we assume that the competitors' prices are not affected by our price decisions. 

Let $H$ \blue{denote the number of subperiods} of one period, and let $h:= 1/H$, e.g., $H=10$ and $h=0.1$. We assume a realization of competitors price trajectories over time via step functions ${\vec p_t}$, $t = 0,h,2h,...,T - h$.




\subsubsection{Deriving Upper Bounds: Evaluation of Forward Looking Strategies}

As we look for upper bounds, in this subsection, we consider the true probabilities $P_t^{(h)}(i,a,\vec p)$, for the subperiods $(t, t+h)$, $t = 0,h,2h,...,T - h$.
If price adjustments are allowed at all points in time $t = 0,h,...,T - h$, the optimal \blue{forward-looking (OFL) pricing policy denoted by $a_t^{OFL(h)}(n;{\vec p_t})$} -- which takes advantage of perfect price anticipations -- is determined by the arg max of, $t = 0,h,...,T-h$, $n = 1,...,N$, $V_T^{OFL(h)}(n;{\vec p_T}):= 0$,
%
\begin{equation}
\label{5.1}
V_t^{OFL(h)}(n;{\vec p_t})
= \mathop {\max }\limits_{a \in A} \left\{ {\sum\limits_{i \ge 0} {P_t^{(h)}(i,a,{{\vec p}_t})}  \cdot \left( {(a - c) \cdot \min (n,i) - n \cdot l \cdot h + {\delta ^h} \cdot V_{t + h}^{OFL(h)}\left( {{{(n - i)}^ + };{{\vec p}_{t + h}}} \right)} \right)} \right\}.
\end{equation}


Next, we consider a second version of the anticipating strategy above, which adjusts prices \textit{less often}.
\blue{Price updates are only allowed at $t = 0,1,...,T - 1$ instead of $t = 0,h,...,T-h$.
To still be able to use a recursive computation (backward induction), we use $a^-$ to store our most recent offer price in an extended state space.} 
\blue{The optimal (forward-looking) pricing policy $a_t^{OFL(1)}(n, a^-;{\vec p_t})$} is determined by the arg max of, $t = 0,h,...,T-h$, ${a^ - } \in A$, $V_T^{OFL(1)}(n,{a^ - };{\vec p_T}):= 0$,
\[V_t^{OFL(1)}(n,{a^ - };{\vec p_t}) = \mathop {\max }\limits_{a \in \left\{ {\begin{array}{*{20}{c}}
A&{,t = 0 \vee t\bmod 1 = 0}\\
{\{ {a^ - }\} }&{,else}
\end{array}} \right.}\]
\begin{equation}
\label{5.2}
\left\{ {\sum\limits_{i \ge 0} {P_t^{(h)}(i,a,{{\vec p}_t})}  \cdot \left( {(a - c) \cdot \min (n,i) - n \cdot l \cdot h + {\delta ^h} \cdot V_{t + h}^{OFL(1)}\left( {{{(n - i)}^ + },a;{{\vec p}_{t + h}}} \right)} \right)} \right\}.
\end{equation}

\smallskip

\blue{Note, the action space in \eqref{5.2} is time-dependent and guarantees that price adjustments are only admissible at $t = 0,1,...,T - 1$ while at all other times the fixed previous price $a^-$ is used.}
In \eqref{5.2} also perfect price anticipations are used.

\blue{Further, using the property} that it is advisable to undercut one of the competitors' prices by the smallest unit $\varepsilon$, cf. Remark \ref{rem4} (ii), we can accelerate the computation of \eqref{5.2} significantly: \blue{(i) instead of $A$, we use the smaller action set} ${\tilde A_t}:= \bigcup\nolimits_{k = 1,...,{K_t}} {\left\{ {{{(p_t^{(k)} - \varepsilon )}^ + }} \right\}} $, $\varepsilon>0$, $t = 0,1,...,T - 1$, and \blue{(ii) we use a smaller state space} by considering only ${a^ - } \in A_t^{( - )}:= {\tilde A_{\left\lfloor {{{(t - h)}^ + }} \right\rfloor }}$, for all points in time $t$, $t = 0,h,...,T$, where $\left\lfloor  \cdot  \right\rfloor $ is the floor operator.




\subsubsection{Evaluation of Non-Anticipating Heuristics}

\blue{Next, we show how to analytically evaluate the non-anticipating policy described in Section 3, cf. Algorithm 3.1}.
First, we allow prices to be adjusted \blue{only at time} $t = 0,1,...,T - 1$. 
The corresponding heuristic strategy, cf.  Algorithm 3.1., is denoted by  ${a^{H(1)}}$.
We consider given \blue{(conditional)} probabilities $\tilde P_t^{(1)}(i,a|\vec p)$ for periods $(t, t+1)$, $t = 0,1,...,T - 1$.
The prices ${a^{H(1)}}(n,{\vec p_t})$ are determined by the arg max of, $t = 0,h,...,T$, $n = 0,1,...,N$, $V_T^{H(1)}(n;{\vec p_T}):= 0$,
\begin{equation}
\label{5.3}
V_t^{H(1)}(n;{\vec p_t}) = \mathop {\max }\limits_{a \in A} \left\{ {\sum\limits_{i \ge 0} {\tilde P_t^{(1)}(i,a|{{\vec p}_t})}  \cdot \left( {(a - c) \cdot \min (n,i) - n \cdot l + \delta  \cdot V_{t + 1}^{H(1)}\left( {{{(n - i)}^ + };{{\vec p}_t}} \right)} \right)} \right\}.
\end{equation}


\blue{Note, \eqref{5.3} defines single prices $a_t^{H(1)}(n,{\vec p_t})$ for time $t$, $t=0,1,...,T-1$, which depend on the current market situation $\vec p_t$ and the current inventory level $n:=X_t$.} 

The expected profit of this non-anticipating heuristic (NAH) policy $a^{H(1)}$ can be \textit{analytically evaluated} via, $t = 0,h,...,T$, $n = 0,1,...,N$, $V_T^{NAH(1)}(n;{\vec p_T}): = 0$,

\[V_t^{NAH(1)}(n;{\vec p_t}) \]
\begin{equation}
\label{5.4}
= \mathop {\max }\limits_{a \in \left\{ {a_{\left\lfloor t \right\rfloor }^{H(1)}(n,{{\vec p}_{\left\lfloor t \right\rfloor }})} \right\}} \left\{ {\sum\limits_{i \ge 0} {P_t^{(h)}(i,a,{{\vec p}_t})}  \cdot \left( {(a - c) \cdot \min (n,i) - n \cdot l \cdot h + {\delta ^h} \cdot V_{t + h}^{NAH(1)}\left( {{{(n - i)}^ + };{{\vec p}_{t + h}}} \right)} \right)} \right\}.
\end{equation}

The recursion \eqref{5.4} is a simple evaluation of the heuristic prices $a_t^{H(1)}(n,{\vec p_t})$, $t = 0,1,...,T - 1$, applied in all subperiods $t = 0,h,...,T$ using the formulation ${a_{\left\lfloor t \right\rfloor }^{H(1)}}(n,{\vec p_{\left\lfloor t \right\rfloor }})$ and the correct probabilities $P_t^{(h)}(i,a,{\vec p_t})$. 
\blue{Note, in (12) no maximization is used as in each state the max operator only evaluates the prices $a = {a_{\left\lfloor t \right\rfloor }^{H(1)}}(n,{\vec p_{\left\lfloor t \right\rfloor }})$ which are uniquely determined by \eqref{5.3}.}

\bigskip
Next, we also consider a second version of the heuristic above, which adjusts prices \textit{more often}. 
Allowing the heuristic, cf. Algorithm 3.1., to adjust prices at all $t = 0,h,...,T - h$, we define the corresponding policy $a_t^{H(h)}(n,{\vec p_t})$ by the arg max of, $t = 0,h,...,T$, $n = 0,1,...,N$, $V_T^{H(h)}(n;{\vec p_T}):= 0$,
\begin{equation}
\label{5.5}
V_t^{H(h)}(n;{\vec p_t}) = \mathop {\max }\limits_{a \in \left\{ {a_t^{H(h)}(n,{{\vec p}_t})} \right\}} \left\{ {\sum\limits_{i \ge 0} {\tilde P_t^{(h)}(i,a|{{\vec p}_t})}  \cdot \left( {(a - c) \cdot \min (n,i) - n \cdot l \cdot h + {\delta ^h} \cdot V_{t + h}^{H(h)}\left( {{{(n - i)}^ + };{{\vec p}_t}} \right)} \right)} \right\}.
\end{equation}
which is a generalization of \eqref{5.3} using \blue{periods of length $h$ instaed of 1 and using $\tilde P_t^{(h)}(i,a|{{\vec p}_t})$ instead of $\tilde P_t^{(1)}(i,a|{{\vec p}_t})$}. An evaluation of the second \blue{fast adjusting} non-anticipating heuristic $a_{}^{H(h)}$ yields the expected profits, $t = 0,h,...,T$, $V_T^{NAH(h)}(n;{\vec p_T}):= 0$,

\[ V_t^{NAH(h)}(n;{\vec p_t})\]
\begin{equation}
\label{5.6}
= \mathop {\max }\limits_{a \in \left\{ {a_t^{H(h)}(n,{{\vec p}_t})} \right\}} \left\{ {\sum\limits_{i \ge 0} {P_t^{(h)}(i,a,{{\vec p}_t})}  \cdot \left( {(a - c) \cdot \min (n,i) - n \cdot l \cdot h + {\delta ^h} \cdot V_{t + h}^{NAH(h)}\left( {{{(n - i)}^ + };{{\vec p}_{t + h}}} \right)} \right)} \right\}.
\end{equation}


\subsubsection{Comparison of Strategies}
Comparing adjustment frequencies and information structures, we obtain relations between the expected profits of \blue{forward-looking strategies, cf. \eqref{5.1}, \eqref{5.2}, and non-anticipating heuristics}, cf. \eqref{5.4}, \eqref{5.6}.

\begin{lemma}
For any given scenario of price trajectories ${\vec p_t}$, $t = 0,h,...,T$, for all $n = 1,...,N$, $t = 0,h,...,T - h$, ${a^ - } \in A$, we have

\medskip
(i)\hspace{0.35cm} $V_t^{OFL(h)}(n) \ge V_t^{OFL(1)}(n,{a^ - }) \ge V_t^{NAH(1)}(n)$,
\medskip

(ii)\hspace{0.29cm} $V_t^{OFL(h)}(n) \ge V_t^{NAH(h)}(n,{a^ - }) \ge V_t^{NAH(1)}(n)$.
\medskip

Proof.	The first inequality of (i) holds since in \eqref{5.1} prices can be adjusted at additional points in time compared to \eqref{5.2}. The second inequality of (i) holds since in \eqref{5.2} the set of admissible prices includes those of \eqref{5.4}. Assertion (ii) follows from similar arguments.
\label{lem1}
\end{lemma}


\blue{Finally, we quantify the performance of our heuristic strategies} compared to the upper bounds derived.

\begin{example}
Consider the setting of Example \ref{ex3} (i)-(iii), $T=100$, $c=3$, $\delta=0.9995$, $N=10$, $d=10$, $l=0.01$, \blue{cf. Figure 4a, 4b, and 4c}. Now, the initial competitors' prices are randomized using ${\vec p_0}: = U(5,15)$. The probability that a single firm $k$ adjusts its \blue{price $p_t^{(k)}$} at $t = 0,h,...,T - h$, is denoted by $\pi \in (0,1)$. We simulate price trajectories using three families of randomized price adjustments, \blue{$k = 1,...,K$, $K=10$}, $t = 0,h,...,T - h$, $h=0.1$, and different \blue{adjustment probabilities} $\pi=0.01,0.03,0.1,0.3$:

\medskip
(i)\hspace{0.35cm} 	no price trend: \hspace{0.9cm} $p_{t + h}^{(k)} = \max \left( {c + 0.01,p_t^{(k)} + {1_{\{ U(0,1) < \pi \} }} \cdot U( - 20,20) \cdot h/\pi /T} \right)$

(ii)\hspace{0.24cm} 	positive price trend: \hspace{0.15cm} $p_{t + h}^{(k)} = \max \left( {c + 0.01,p_t^{(k)} + {1_{\{ U(0,1) < \pi \} }} \cdot U( - 15,25) \cdot h/\pi /T} \right)$

(iii)\hspace{0.12cm} 	negative price trend:  \hspace{0.07cm} $p_{t + h}^{(k)} = \max \left( {c + 0.01,p_t^{(k)} + {1_{\{ U(0,1) < \pi \} }} \cdot U( - 25,15) \cdot h/\pi /T} \right)$

\medskip
In the three settings of Example \ref{ex4}, we apply the following five different strategies, cf. \eqref{5.1}-\eqref{5.6}:
\smallskip

(A)\hspace{0.17cm}	Frequent informed:\hspace{0.4cm}	adjustments in $t = 0,h,...,T - h$,	full anticipation, cf. ${V^{OFL(h)}}$
  
(B)\hspace{0.2cm}	Relaxed informed:\hspace{0.52cm}	adjustments in $t = 0,1,...,T - 1$,	full anticipation, cf. ${V^{OFL(1)}}$
  
(C)\hspace{0.2cm}	Frequent heuristic:\hspace{0.4cm}	adjustments in $t = 0,h,...,T - h$,	no anticipation, cf. ${V^{NAH(h)}}$
  
(D)\hspace{0.2cm}	Relaxed heuristic:\hspace{0.52cm}	adjustments in $t = 0,1,...,T - 1$,	no anticipation, cf. ${V^{NAH(1)}}$
  
(E)\hspace{0.2cm}	Optimal fix price:\hspace{0.6cm}		adjustment only in $t = 0$,		full anticipation 

\medskip

\noindent We use $P_t^{(h)}(i,a,\vec p):= Pois( {h \cdot d \cdot \hat P(a,\vec p)})$
and $\tilde P_t^{(\Delta)}(i,a|\vec p):= Pois( {\Delta \cdot d \cdot \hat P(a,\vec p)}$, 
$\Delta \in \{h,1\}$. 
\label{ex4}
\end{example}

\smallskip

A comparison of the performance of the different strategies (A) - (E) applied in the \blue{three} settings (i) - (iii) of Example \ref{ex4} are summarized in Table 1. \blue{For different competitors' price adjustment frequencies $\pi$} the table contains the average value of the expected profits $V_0^{(A)}(N;{\vec p_0})$ of the benchmark strategy A \blue{for various random scenarios ${\vec p_t}$}. The expected profit $V_0^{(A)}(N;{\vec p_0})$ represents an upper bound as full information of future prices is used and the possibility of price adjustments is the largest. The remaining columns of the table contain the average ratio of the expected profits of the strategies (B) - (E) compared to strategy (A). For each simulated competitive scenario the expected profits of all five strategies have been evaluated analytically. For each setting (i) - (iii) 1000 scenarios of \blue{trajectories $\vec p_t$, $t = 0,h,...,T - h$, were evaluated.}



\begin{table}[ht]
\begin{center}
\setlength{\tabcolsep}{4mm}
\begin{tabular}{c ccc ccc}
\toprule
  ${\vec p_t}$     &	$\pi$	&$V_0^{(A)}(N;{\vec p_0})$		
&$\frac{{V_0^{(B)}(N;{{\vec p}_0})}}{{V_0^{(A)}(N;{{\vec p}_0})}}$			
&$\frac{{V_0^{(C)}(N;{{\vec p}_0})}}{{V_0^{(A)}(N;{{\vec p}_0})}}$			
&$\frac{{V_0^{(D)}(N;{{\vec p}_0})}}{{V_0^{(A)}(N;{{\vec p}_0})}}$			
&$\frac{{V_0^{(E)}(N;{{\vec p}_0})}}{{V_0^{(A)}(N;{{\vec p}_0})}}$	\\ \midrule
(i)	&0.01	&22.53	&0.983	&0.986	&0.964	&0.706  \\
(i)	&0.03 	&25.31	&0.973	&0.985	&0.953	&0.760  \\
(i)	&0.10	&26.73	&0.948	&0.987	&0.926	&0.802  \\
(i)	&0.30	&26.87	&0.911	&0.990	&0.884	&0.836  \\ \midrule
(ii)	&0.01	&36.11	&0.990	&0.964	&0.954	&0.781  \\
(ii)	&0.03	&41.50	&0.983	&0.943	&0.930	&0.786  \\
(ii)	&0.10	&43.96	&0.968	&0.932	&0.910	&0.780  \\
(ii)	&0.30	&45.08	&0.953	&0.926	&0.898	&0.772  \\ \midrule
(iii)	&0.01	&10.83	&0.959	&0.976	&0.920	&0.345  \\
(iii)	&0.03	&12.04	&0.930	&0.984	&0.891	&0.440  \\
(iii)	&0.10	&12.51	&0.844	&0.986	&0.793	&0.469  \\
(iii)	&0.30	&12.37	&0.686	&0.984	&0.629	&0.474  \\
\bottomrule
\end{tabular}
\caption{Expected profits of the heuristic strategy compared to upper bounds for simulated competitive scenarios (i)-(iii), $T=100$, $N=10$, $d=10$, $h=0.1$, and $\pi=0.01,0.03,0.1,0.3$; Example \ref{ex4}.
\label{tab1}}
\end{center}
\end{table}


The relaxed heuristic strategy (D) yields 63-96\% of the optimal results; the frequent heuristic strategy (C) even obtains 93-99\%. 
The performance of the strategies (B) and (C) is in between those of strategy (A) and (D), cf. Lemma \ref{lem1}.
We observe that the more volatile the market, the more important become adjustment frequencies. 
Our example shows that the frequent heuristic strategy (C) can even beat the relaxed informed strategy (B).
\blue{The best \textit{fix price} strategy (E)} only yields 35-84\% of the optimal profits, although future prices are fully known.
A boxplot of Table 1 is given in the Appendix, see Figure A.10.  

Note, while \blue{in Example 5.1 $P_t^{(h)}$ and $\tilde P_t^{(h)}$ coincide (which in real-life applications can be justified for small intervals $h$)}, the probabilities $\tilde P_t^{(1)}$ used in \eqref{5.3} might strongly over- or underestimate the correct conditional probabilities in the specific settings of Example 5.1 (i) - (iii). Hence, results of strategy (D) could be even improved by estimating $\tilde P_t^{(1)}$ accurately.

In further simulations, we also varied other parameters, such as time horizon $T$, initial items $N$, number of competitors $K$, as well as the demand parameters $\beta$ or the adjustment dynamics of the competitors. Overall, the results were similar. We summarize the most \blue{important findings} 
in the following remark.

\begin{remark}	(Impact of adjustment frequencies and price anticipations)   

(i)\hspace{0.41cm} In general, adjustment frequencies as well as price anticipations significantly raise profits.

(ii)\hspace{0.29cm} Higher adjustment frequencies, cf. heuristic (C), can overcompensate the value of price 

\hspace{0.95cm}anticipations, cf. strategy (B).

(iii)\hspace{0.2cm} The performance of fix price strategies is low, even if they are optimally chosen and perfect price 

\hspace{0.95cm}anticipations are taken into account.

(iv)\hspace{0.3cm} If market prices have a positive trend, cf. scenario (ii), then price anticipations are more important 

\hspace{0.95cm}than frequent price adjustments.

(v)\hspace{0.38cm} If market prices have a negative trend, cf. scenario (iii), then frequent price adjustments are more 

\hspace{0.95cm}important than price anticipations.

(vi)\hspace{0.26cm} If prices are adjusted frequently, then price anticipations are less important.

(vii)\hspace{0.14cm} If market volatility is high, then it is important to adjust prices frequently.

\label{rem5}
\end{remark}



\subsection{Heuristic vs. Optimal Response Strategies in Strategic Duopoly Competition}

\blue{In this subsection, we consider strategic competitors, who choose his/her prices in response to our current price. The goal is to} compare our heuristic approach to optimal response strategies. Since optimal response strategies cannot be computed for complex competition settings (due to the curse of dimensionality), we consider a duopoly setting.


\subsubsection{Optimal Response Strategies \blue{in a Duopoly}}

We want to compare the performance of our heuristic strategy derived in Section 3 to upper bounds, which are determined by optimal response strategies that make use of full information. We assume that the price reaction of the competitor as well as the reaction times can be fully anticipated. Note, our heuristic strategy does not use this additional information, which corresponds to real-life scenarios.

\blue{We assume that the market situation (state) is one-dimensional} and simply characterized by the competitor's price $p$, i.e., we let $\vec s:= p$. We assume that the competitor adjusts its price $p$ in response to \blue{our price $a$} with a fixed delay of $\Delta$ periods, $0 < \Delta  < 1$. Choosing a price $a$ at time $t$ is followed by the competitor's price reaction $F$, which can depend on $p$ and $a$. I.e., after an interval of size $\Delta$ the competitor adjusts its price from $p$ to $F(a,p)$. \blue{Consequently, our reaction time is $1-\Delta$}.

Following the examples of the previous sections, in period $t$ the probability to sell exactly $i$ items during the first interval of size $\Delta$ (Phase 1) is $P_t^{(\Delta )}(i,a,p):= Pois( {\Delta  \cdot d \cdot \hat P(a,p)} )$. For the rest of the period (Phase 2) the sales probability changes to $P_{t + \Delta }^{(1 - \Delta )}\left( {i,a,F(a,p)} \right) = Pois( {(1 - \Delta ) \cdot d \cdot \hat P\left( {a,F(a,p)} \right)} )$. 

In case the competitors' strategy is known the Hamilton-Jacobi-Bellman (HJB) equation of \blue{the duopoly problem} can be written as, $t = 0,1,...,T - 1$, $n = 1,...,N$, $p \in A$, $0 < \Delta  < 1$,
%

\[{V_t}^{(\Delta)*}(n,p) = \mathop {\max }\limits_{a \in A} \left\{ {\sum\limits_{{i_1} \ge 0} {P_t^{(\Delta )}({i_1},a,p)}  \cdot \sum\limits_{{i_2} \ge 0} {P_{t + \Delta }^{(1 - \Delta )}\left( {{i_2},a,F(a,p)} \right)} } \right.\]
\begin{equation}
\label{5.7}
\left. { \cdot \left( {(a - c) \cdot \min (n,{i_1} + {i_2}) - n \cdot l + \delta  \cdot V_{t + 1}^{(\Delta )*}\left( {{{(n - {i_1} - {i_2})}^ + },F(a,p)} \right)} \right)} \right\},
\end{equation}
%
where 
$V_T^{(\Delta )*}(n,p) = 0$ for all $n,p$. The associated optimal pricing strategy $a_t^{(\Delta )*}(n,p)$, $t = 0,1,...,T - 1$, $n = 1,...,N$, $p \in A$, is determined by the arg max of \eqref{5.7}.



\subsubsection{Heuristic Response Strategies \blue{in a Duopoly}}

\blue{In case neither the competitor's strategy nor his/her reaction time $\Delta$ is known, our heuristic strategy, cf. Algorithm 3.1, can be applied using given conditional probabilities ${\tilde P_t^{(1)}(i,a|p)}$.}
The corresponding strategy that adjusts prices at $t=0,1,...,T-1$ \blue{denoted by} $\tilde a_t^{(\Delta )}(n,p)$, $n = 1,...,N$, $p \in A$, is defined by the arg max of, $t = 0,1,...,T - 1$, $n = 1,...,N$, $p \in A$, $0 < \Delta  < 1$, $\tilde V_T^{(\Delta )}(n,p):= 0$ for all $n,p$,

\begin{equation}
\label{5.8}
\tilde V_t^{(\Delta )}(n,p) = \mathop {\max }\limits_{a \in A} \left\{ {\sum\limits_{i \ge 0}{\tilde P_t^{(1)}(i,a|p)}  \cdot \left( {(a - c) \cdot \min (n,i) - n \cdot l + \delta  \cdot \tilde V_{t + 1}^{(\Delta )}\left( {{{(n - i)}^ + },p} \right)} \right)} \right\}.
\end{equation}

The performance of the heuristic strategy $\tilde a_t^{(\Delta )}$, cf. \eqref{5.8}, can \blue{again} be evaluated analytically and yields the (suboptimal) expected profits, $t = 0,1,...,T - 1$, $n = 0,1,...,N$, $p \in A$, \blue{$0 < \Delta  < 1$}, ${\bar V_T}^{(\Delta )}(n,p) = 0$,

\[{\bar V_t}^{(\Delta )}(n,p) = \mathop {\max }\limits_{a \in \{ \tilde a_t^{(\Delta )}(n,p)\} } \left\{ {\sum\limits_{{i_1} \ge 0} {P_t^{(\Delta )}\left( {{i_1},a,p} \right)}  \cdot \sum\limits_{{i_2} \ge 0} {P_{t + \Delta }^{(1 - \Delta )}\left( {{i_2},a,F\left( {a,p} \right)} \right)} } \right.\]
\begin{equation}
\label{5.9}
\left. { \cdot \left( {(a - c) \cdot \min (n,{i_1} + {i_2}) - n \cdot l + \delta  \cdot \bar V_{t + 1}^{(\Delta )}\left( {{{(n - {i_1} - {i_2})}^ + },F\left( {a,p} \right)} \right)} \right)} \right\}.
\end{equation}


Note, due to the reaction of the competitor, the probabilities $P_t^{(1)}(i,a,p)$ and $\tilde P_t^{(1)}(i,a|p)$ can be quite \textit{different}, e.g., if the competitor quickly undercuts our price and our price rank changes, \blue{cf. Figure 1a}. Hence, assuming stable prices and using $\tilde P_t^{(1)}(i,a|p):= P_t^{(1)}(i,a,p)$ might (strongly) overestimate the conditional sales probabilities.
Hence, \blue{in such a case, observed sales will indicate that sales probabilities $\tilde P_t^{(1)}(i,a|p)$} significantly differ from $P_t^{(1)}(i,a,p)$.

In general, the characteristics of a specific competitive setup will be reflected in sales data. Analyzing the relation between realized sales, offer prices, and the underlying market situations that have been observed at the time of the price adjustment, cf. Section 4.1, makes it possible to estimate the conditional sales probabilities $\tilde P_t^{(1)}(i,a|p)$, see also, e.g., 
Vulcano et al. (2012), Abdallah, Vulcano (2016), and Fisher et al. (2017).
Note, neither the strategy nor the reaction time of the competitor has to be discovered.
An accurate estimation of $\tilde P_t^{(1)}(i,a|p)$ may provide 

\begin{equation}
\label{5.10}
\tilde P_t^{(1)}(i,a|p) \approx \sum\limits_{{i_1},{i_2} \ge 0{:_{}}{i_1} + {i_2} = i} {P_t^{(\Delta )}\left( {{i_1},a,p} \right) \cdot P_{t + \Delta }^{(1 - \Delta )}\left( {{i_2},a,F(a,p)} \right)}.
\end{equation}


\subsubsection{Comparison of Strategies \blue{in a Duopoly}}

Next, we want to measure the performance of the heuristic compared to the \blue{optimal (informed) strategy}. 
The following example studies one of the most common strategies \blue{which always slightly undercuts} the competitor's price, \blue{cf. Example 4.3 (iv), Fig. 4d}.
We also study impact of the reaction time $\Delta$ as well as the accuracy of the estimation of $\tilde P_t^{(1)}$.

\smallskip

\begin{example}
We assume a duopoly, i.e., $K=1$. Let $T=100$, $N=5$, $l=0.01$, $c=3$, $\delta  = 0.9995$, $0<\Delta<1$, $d=10$. The demand is defined as in Example \ref{ex3}, i.e., $P_t^{(\Delta )}(i,a,p):= Pois( {\Delta  \cdot d \cdot \hat P(a,p)} )$. We let our competitor play the response strategy $F(a,p):= \max (a - \varepsilon ,c)$, $\varepsilon=1$, $A:= \left\{ {1,2,...,120} \right\}$. The competitor adjusts prices in $t = \Delta ,1 + \Delta ,...,T - 1 + \Delta$. We adjust prices at $t = 0,1,...,T - 1$. We simulate three different \blue{price adjustment} strategies:
\smallskip

(i)\hspace{0.4cm} the optimal response strategy $^{(i)}a_{}^{(\Delta )*}$, cf. \eqref{5.7},

(ii)\hspace{0.3cm} the heuristic strategy ${^{(ii)}\tilde a^{(\Delta)}}$, cf. \eqref{5.8}, based on $\tilde P_t^{(1)}(i,a|p):= P_t^{(1)}(i,a,p)$, and

(iii)\hspace{0.2cm} the heuristic strategy ${^{(iii)}\tilde a^{(\Delta)}}$, cf. \eqref{5.8}, based on \blue{(accurate)} $\tilde P_t^{(1)}(i,a|p)$ defined as in \eqref{5.10}.
\label{ex5}
\end{example}

\smallskip

Figure 5 shows optimal response strategies \blue{for Example 5.2 (i) - (iii) at time} $t=0$ for different inventory levels $n$ and competitor's prices $p$ \blue{in case of} $\Delta=0.5$. Figure 5a illustrates the optimal policy; Figure 5b and 5c depict the heuristic policy in case (ii) and (iii), respectively. \blue{The optimal response strategy (i) is of similar shape for different inventory levels:} If the competitor's price is very low, it is optimal to raise the price up to a certain price level. Almost the same price level has to be chosen if the competitor's price is very high. If the competitor's price is somewhere in between (intermediate range), it is best to undercut that price by one price unit $\varepsilon$. If $n$ increases, then the upper price level decreases; the intermediate range is of the same size but at a lower level.



\begin{figure}[ht]
\begin{center}
\includegraphics[scale=0.45]{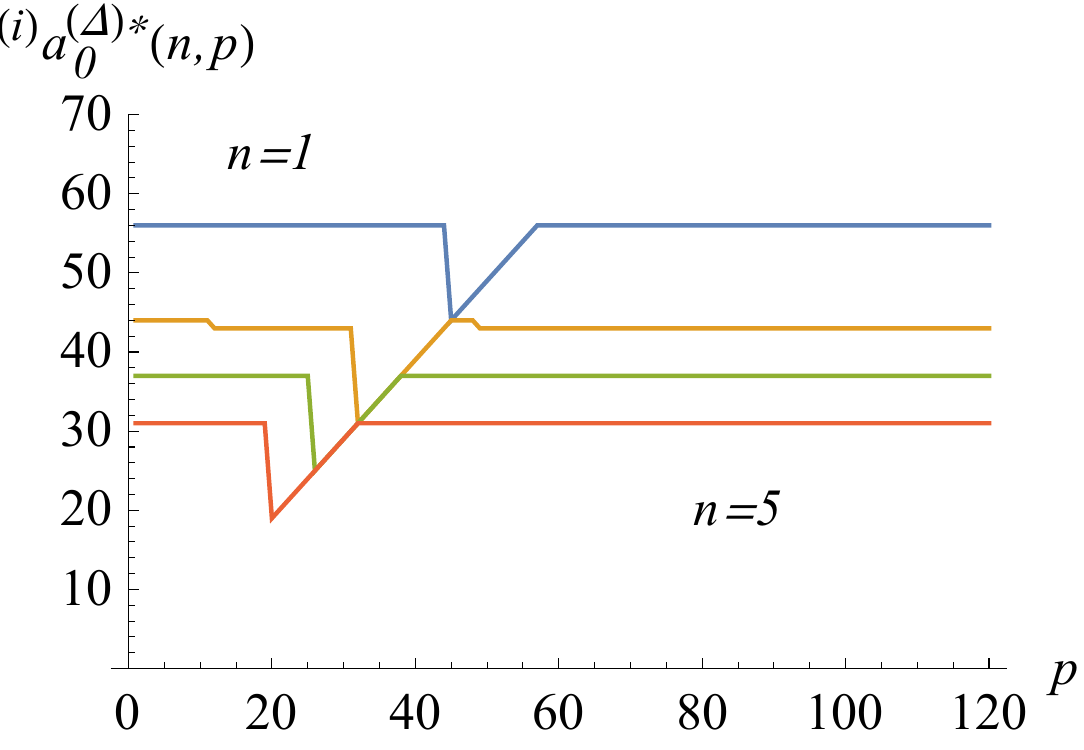}\quad
\includegraphics[scale=0.45]{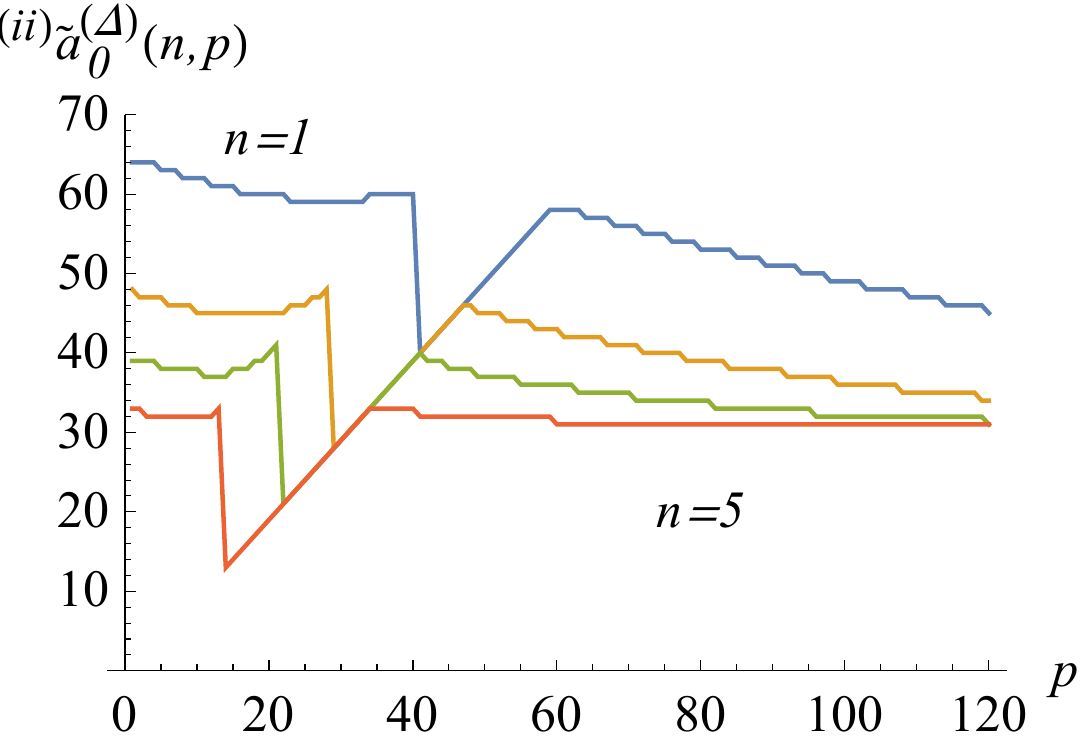}\quad
\includegraphics[scale=0.45]{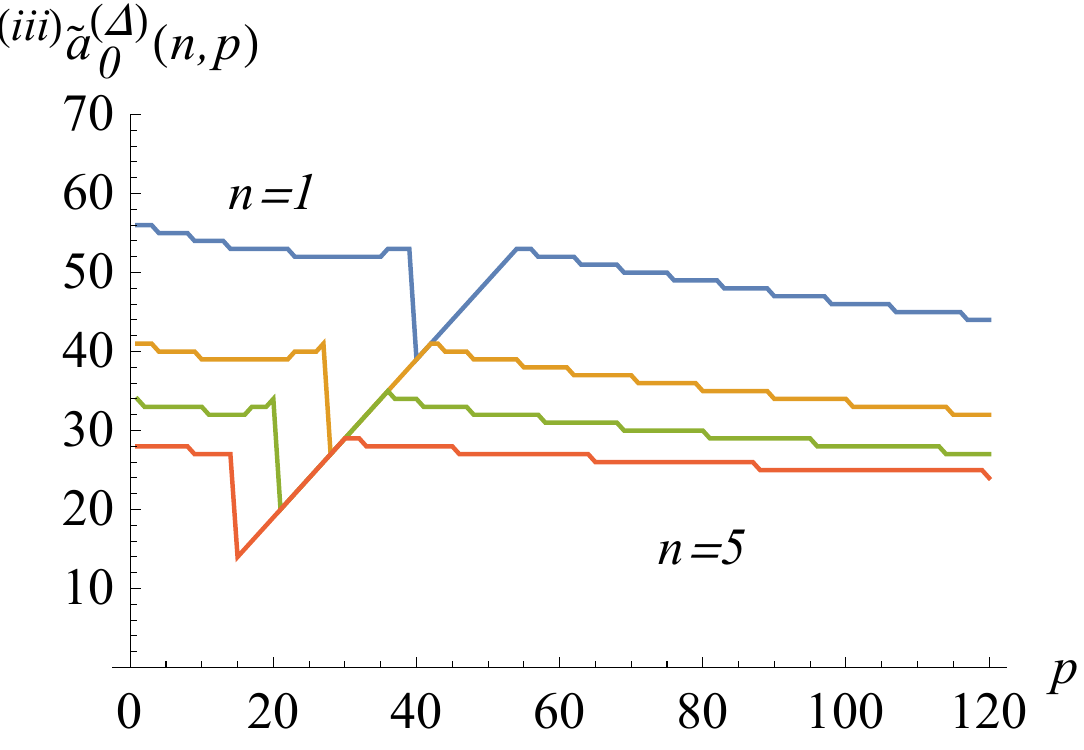}
\caption{Comparison of feedback prices \blue{for different inventory levels $n$}: Optimal response policy ${^{(i)}a_0^{(\Delta)*}}(n,p)$ (left window 5a) and heuristic policy in case (ii) $^{(ii)}\tilde a_0^{(\Delta )}(n,p)$ (middle window 5b), and case (iii) $^{(iii)}\tilde a_0^{(\Delta )}(n,p)$  (right window 5c) for 
$n=1, 2, 3, 5$, $\Delta=0.5$; Example \ref{ex5}.}
\label{fig5}
\end{center}
\end{figure}



\begin{figure}[ht]
\begin{center}
\includegraphics[scale=0.45]{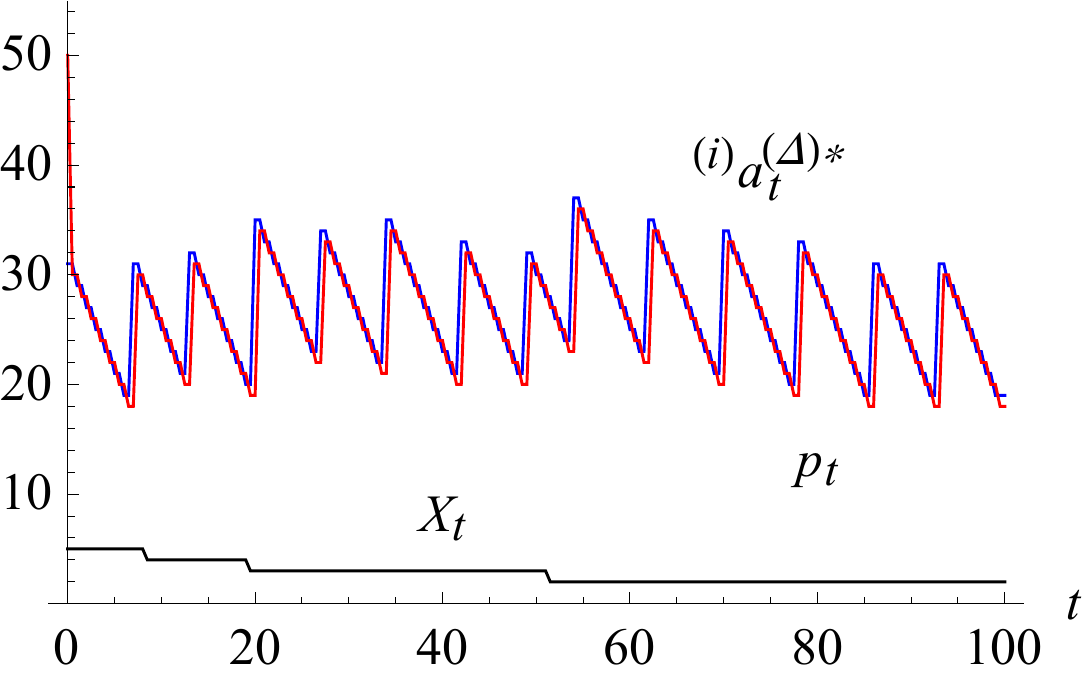}\quad
\includegraphics[scale=0.45]{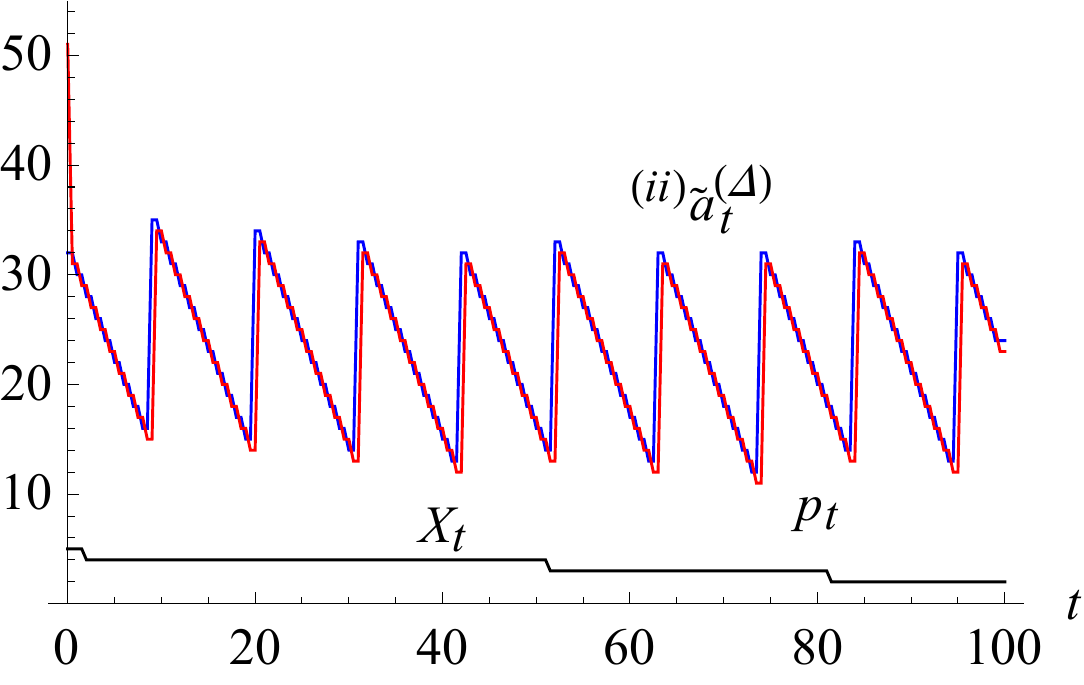}\quad
\includegraphics[scale=0.45]{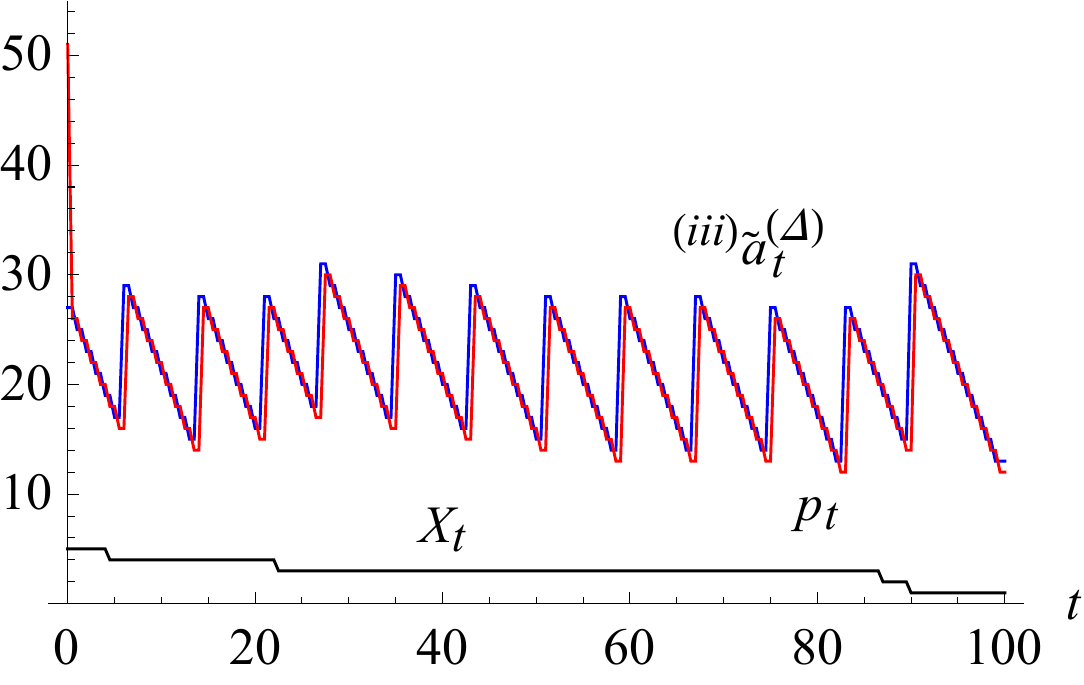}
\caption{Comparison of simulated price paths over time: Optimal policy ${^{(i)}a_0^{(\Delta)*}}(n,p)$ (left window 6a) and heuristic policy in case (ii) $^{(ii)}\tilde a_0^{(\Delta )}(n,p)$ (middle window 6b), and case (iii) $^{(iii)}\tilde a_0^{(\Delta )}(n,p)$ (right window 6c) for the case $N=5$, $T=100$, $\Delta=0.5$; Example \ref{ex5}.}
\label{fig6}
\end{center}
\end{figure}

\blue{Overall, the heuristic policies have similar characteristics.}
Compared to the optimal response policy and heuristic (iii), the prices of heuristic (ii) are lower and the intermediate range is wider. \blue{The corresponding figures of case (i) and (iii) for $\Delta=0.1$ and $\Delta=0.9$ are shown in the appendix, cf. Figure A.11 and A.12}. Note, the policy of case (ii) is independent of $\Delta$ \blue{as the reaction time is neither explicitly nor implicitly taken into account}. We observe that the intermediate range of the optimal policy is increasing in $\Delta$. Policy (i) and (iii) are quite similar for all $\Delta$ values. If $\Delta$ is large then policy (ii) is similar to the other two. If $\Delta$ is small then policy (ii) differs significantly.

The application of \blue{the optimal and both heuristic} strategies leads to cyclic price patterns over time, cf. Edgeworth cycles, see, e.g., Maskin, Tirole (1988) or Noel (2007). The resulting price paths for the optimal response strategy and the heuristic strategies are shown in Figure 6. We observe that compared to heuristic (ii) the cycle length and the amplitude of the price patterns are smaller for the optimal policy (i) and the heuristic (iii). \blue{Finally, we obtain that heuristic (iii) is very similar to the optimal strategy (i).}

The cyclic \blue{price patterns} are basically independent of the initial price. Hence, if the time horizon and the discounting factor are not too "small" then the expected future long-term profits ${V^{(\Delta)*}}(n,p)$ and ${\bar V_0^{(\Delta)}}(n,p)$, respectively, are (almost) the same for different initial prices $p$ (in our example the maximum deviation is 0.1). The number of items left to sell and particularly the reaction time $\Delta$ plays a more prominent role.

\medskip


\begin{table}[ht]
\begin{center}
\setlength{\tabcolsep}{1mm}
\begin{tabular}{c ccc ccc}
\toprule
$n$
&$^{(i)}{V_0}^{(0.1)*}(n,50)$	
&$\frac{{{}^{(ii)}{{\bar V}_0}^{(0.1)}(n,50)}}{{^{(i)}{V_0}^{(0.1)*}(n,50)}}$	
&$\frac{{{}^{(iii)}{{\bar V}_0}^{(0.1)}(n,50)}}{{^{(i)}{V_0}^{(0.1)*}(n,50)}}$
&$^{(i)}{V_0}^{(0.9)*}(n,50)$	
&$\frac{{{}^{(ii)}{{\bar V}_0}^{(0.9)}(n,50)}}{{^{(i)}{V_0}^{(0.9)*}(n,50)}}$	
&$\frac{{{}^{(iii)}{{\bar V}_0}^{(0.9)}(n,50)}}{{^{(i)}{V_0}^{(0.9)*}(n,50)}}$\\ \midrule
1	&23.3637	&0.9801	&0.9949	&29.0480	&0.9881	&0.9852  \\
2	&34.5616	&0.9766	&0.9942	&45.2496	&0.9867	&0.9841  \\
3	&39.7475	&0.9716	&0.9925	&54.4413	&0.9801	&0.9803  \\
5	&41.9375	&0.9584	&0.9910	&61.5614	&0.9731	&0.9761  \\
7	&40.6005	&0.9473	&0.9890	&61.9205	&0.9690	&0.9774  \\
10	&37.7302	&0.9413	&0.9879	&59.4264	&0.9675	&0.9795  \\ \bottomrule
\end{tabular}
\caption{Comparison of expected profits for \blue{different estimations of sales probabilities:
Optimal policy ${^{(i)}a^{(\Delta)*}}$, cf. (i), vs. heuristic policies} ${^{(ii)}\tilde a^{(\Delta)}}$, cf. (ii), and ${^{(iii)}\tilde a^{(\Delta)}}$, cf. (iii), $\Delta=0.1,0.9$, $n=1,2,3,5,7,10$; Example \ref{ex5}.
\label{tab2}}
\end{center}
\end{table}

In Table 2, the expected profits $^{(i)}{V^{(\Delta )*}}$, $^{(ii)}{\bar V^{(\Delta )}}$, and $^{(iii)}{\bar V^{(\Delta )}}$ of Example \ref{ex5} are compared for $\Delta=0.1$ and $\Delta=0.9$. 
In our example, the expected profits $V^{(\Delta )*}$ and $\bar V^{(\Delta )}$ are maximized for an inventory level of 5-7 items. A higher number of items is not advantageous which is due to inventory holding costs. 


\blue{Finally, we observe} that the performance of our heuristic is quite promising. Although the heuristic does not anticipate the competitor's \blue{blue} reaction, in case (ii), 94-98\% of the theoretically optimal expected profits are obtained; in case (iii), we obtain even 98-99\%. 

\blue{If the competitor's reaction time $\Delta$ is small, the results of heuristic (iii) outperform} those of heuristic (ii) by up to 5\%. The reason is the following: If $\Delta$ is large then the market situation hardly changes during one period of time and, \blue{thus}, $P_{}^{(1)}$ and $\tilde P_{}^{(1)}$ are almost identical. If $\Delta$ is \blue{small} then the market situation quickly changes, i.e., $P_{}^{(1)}$ and $\tilde P_{}^{(1)}$ differ significantly, and in turn, the performance of the heuristic suffers.


\subsubsection{Comparison of Reaction Times \blue{in a Duopoly}}

Due to the significant impact of reaction times, firms will try to minimize their reaction times by anticipating their competitors' time of adjustment and to time their price adjustments optimally. In order not to act predictably, in turn, strategic firms might randomize their reaction times.
\blue{Finally, the} percentage of time a firm has the most recent price will be determined by the ratio of \blue{the firms'} adjustment frequencies. 
\blue{Refreshing the prices $x$ times as often as the competitor can be compared to our model with $\Delta:= x/(1 + x)$, or $x = \Delta/(1 - \Delta)$, $x > 0$.} For instance, the ratio $x=2.33$ corresponds to a value of $\Delta=0.7$. Note, while some firms adjust their prices once a day, other firms adjust their prices every one or two hours.

In the following, we study to which extent expected profits of our strategies are affected by $\Delta$. Following Example \ref{ex5}, we compare the expected profits ${V^{(\Delta)*}}$ of the optimal policy and ${\bar V^{(\Delta)}}$ \blue{of the heuristic policy with the benchmark profit} ${V^{(0.5)*}}$  (for the case $\Delta=0.5$ or $x=1$, respectively).



\begin{table}[ht]
\begin{center}
\setlength{\tabcolsep}{2mm}
\begin{tabular}{cccccccc}
\toprule
Strategy	 & \blue{Competitor's reaction time $\Delta:$}	&0.1	&0.3	&0.5	&0.55	&0.7	&0.9  \\ \hline
(i)	 &${V_0}^{(\Delta )*}(1,50)/{V_0}^{(0.5)*}(1,50)$
&0.8873	&0.9444	&1.0000	&1.0135	&1.0529	&1.1032 \\
(i)	 &${V_0}^{(\Delta )*}(5,50)/{V_0}^{(0.5)*}(5,50)$
&0.8101	&0.9041	&1.0000	&1.0239	&1.0954	&1.1892 \\
(i)	 &${V_0}^{(\Delta )*}(10,50)/{V_0}^{(0.5)*}(10,50)$
&0.7799	&0.8878	&1.0000	&1.0284	&1.1138	&1.2284 \\ \midrule
(ii)	 &${\bar V_0}^{(\Delta )}(1,50)/{V_0}^{(0.5)*}(1,50)$
&0.8697	&0.9333	&0.9908	&1.0043	&1.0429	&1.0900 \\
(ii)	 &${\bar V_0}^{(\Delta )}(5,50)/{V_0}^{(0.5)*}(5,50)$
&0.7765	&0.8762	&0.9730	&0.9968	&1.0669	&1.1573 \\
(ii)	 &${\bar V_0}^{(\Delta )}(10,50)/{V_0}^{(0.5)*}(10,50)$
&0.7341	&0.8478	&0.9614	&0.9898	&1.0750	&1.1884 \\ \midrule
(iii)	 &${\bar V_0}^{(\Delta )}(1,50)/{V_0}^{(0.5)*}(1,50)$
&0.8828	&0.9331	&0.9882	&1.0005	&1.0370	&1.0868 \\
(iii)	 &${\bar V_0}^{(\Delta )}(5,50)/{V_0}^{(0.5)*}(5,50)$
&0.8028	&0.8858	&0.9710	&0.9988	&1.0650	&1.1601 \\
(iii)	 &${\bar V_0}^{(\Delta )}(10,50)/{V_0}^{(0.5)*}(10,50)$
&0.7705	&0.8697	&0.9722	&1.0024	&1.0838	&1.2032 \\ \bottomrule
\end{tabular}
\caption{Additional information vs. reaction time: Optimal policy ${a^{(\Delta)*}}$, cf. (i), and heuristic policies ${\tilde a^{(\Delta )}}$, cf. (ii) and (iii), for different \blue{competitor's} reaction times $0.1 \le \Delta  \le 0.9$ compared to the \blue{expected benchmark profits ${V_0^{(0.5)*}}$ of policy ${a^{(0.5)*}}$ in case $\Delta =0.5$}, $n=1,5,10$; Example \ref{ex5}.
\label{tab3}}
\end{center}
\end{table}



Table 3 illustrates the ratio ${V^{(\Delta)*}}/{V^{(0.5)*}}$ as well as the ratio ${\bar V^{(\Delta)}}/{V^{(0.5)*}}$ for different $n$ and $\Delta$ in case (i) - (iii) of Example \ref{ex5}.
We observe that the reaction time $\Delta$ has a significant impact on expected profits. \blue{On the one hand,} our example shows that small reaction frequencies can shrink expected profits by more than 20\%, cf. $\Delta$=0.1. \blue{On the other hand,} heuristics using shorter reaction times or more frequent price adjustments, respectively, can easily overcompensate a lack of information, cf. $\Delta$=0.55, and increase expected profits by more than 20\%, cf. $\Delta$=0.9.
We conclude that the more volatile the market is, the more important is it to use an accurate estimation of conditional probabilities $\tilde P$, which are also characterized by the own adjustment frequency. Note, it is not necessary to know the competitors' reaction \blue{strategy -- the information} of the competitors' behavior and the market dynamics are implicitly contained in $\tilde P$.

\smallskip

\begin{remark}  (Main insights)                 

(i)\hspace{0.4cm}	For successful strategies regular price adjustments as well as anticipation of market dynamics are important. Both effects are challenging: on the one hand, computations of dynamic systems are time-consuming; on the other hand, real-life market dynamics are usually complex and characterized by a huge number of factors.

(ii)\hspace{0.3cm}	To circumvent the curse of dimensionality, we \blue{simplify the problem: 
While the reduced state space} of "time and inventory level" allows for fast optimization, in our model the anticipation of market evolutions is transferred to the model's 
sales probabilities. The inaccuracy of our simplifying "stability assumption" -- which is the key for the decomposition -- can be compensated as frequent \blue{observations of the current market situation and corresponding price adjustments keep the system} up to date.


(iii)\hspace{0.2cm}	In real-life applications, the conditional sales probabilities of the model can be estimated from market data using state-of-the-art regression models (logistic regression, decision trees, neural networks, etc.).
\label{rem6}
\end{remark}

\medskip

Our examples verify that the performance of our heuristic compared to optimal policies (that use full information) is excellent even if strategic competitors are involved. While optimal solutions are limited to very few competitors, the heuristic strategy instead can easily be applied in the presence of even very large numbers of competitors. Since the state space of the heuristic approach just consists of two dimensions, the computation of prices is constantly fast. Hence, it is ensured that it is possible to compute prices for complex situations and to adjust prices with high frequencies.



\subsection{Heuristic vs. Heuristic Response Strategies in Strategic Oligopoly Competition}

In this subsection, we study the case in which \blue{our heuristic strategy} is played by multiple firms against each other.
%
In the next example, we consider such a scenario. 

\smallskip

\begin{example}
Consider the setting of Example \ref{ex4} (i). Let $K=3$, $T=100$, $N=10$, $l=0.01$, $c=3$, $\delta  = 0.9995$, $d=40$, and $A:= \left\{ {1,2,...,100} \right\}$. All competitors play the heuristic strategy, cf. Algorithm 3.1. 
All firms use randomized reaction times: In every point in time $t = 0,h,...,T - h$ prices are adjusted by each player with probability $\pi  = h$, $h=0.1$. Initial prices are given by $p_0^{(k)}:= 20$, $k = 1,...,K$. For evaluation we use the sales probabilities $P_t^{(h)}(i,a,\vec p):= Pois( {h \cdot d \cdot \hat P(a,\vec p)} )$. For computation of the heuristics, all firms use $\tilde P_t^{(1)}(i,a|\vec p): = P_t^{(1)}(i,a,\vec p)/K$ as a simple estimation for the conditional probabilities.
\label{ex6}
\end{example}

\medskip


\begin{figure}[ht]
\begin{center}
\includegraphics[scale=0.63]{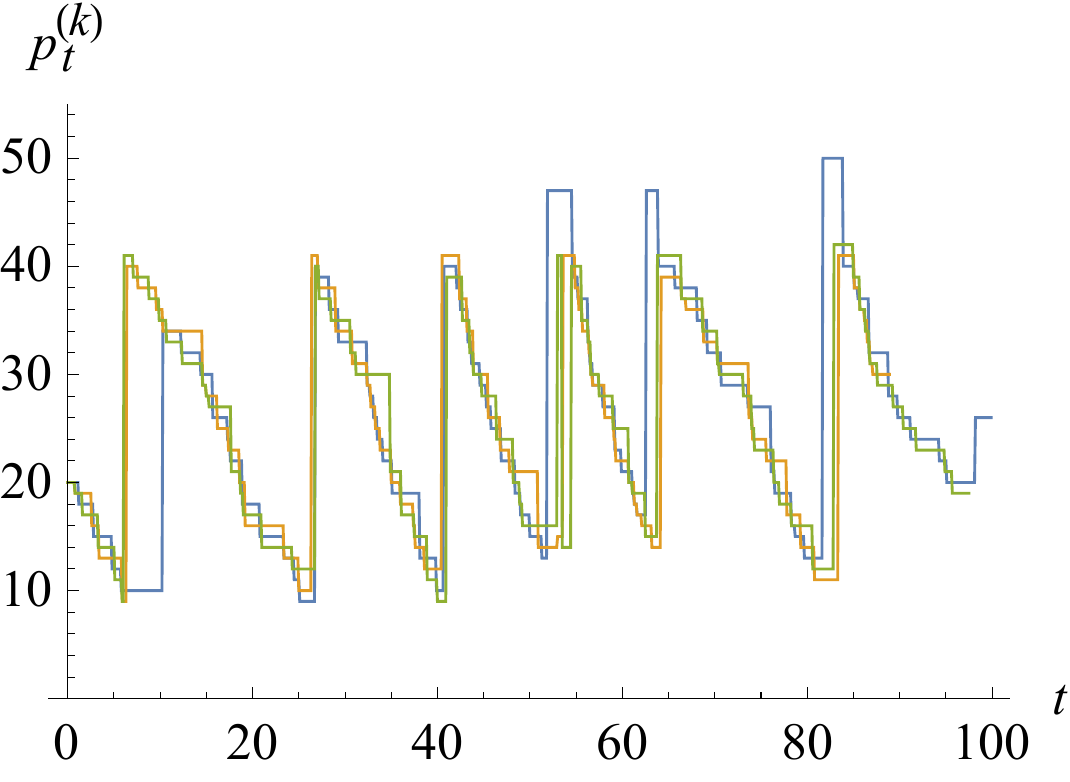}\qquad
\includegraphics[scale=0.63]{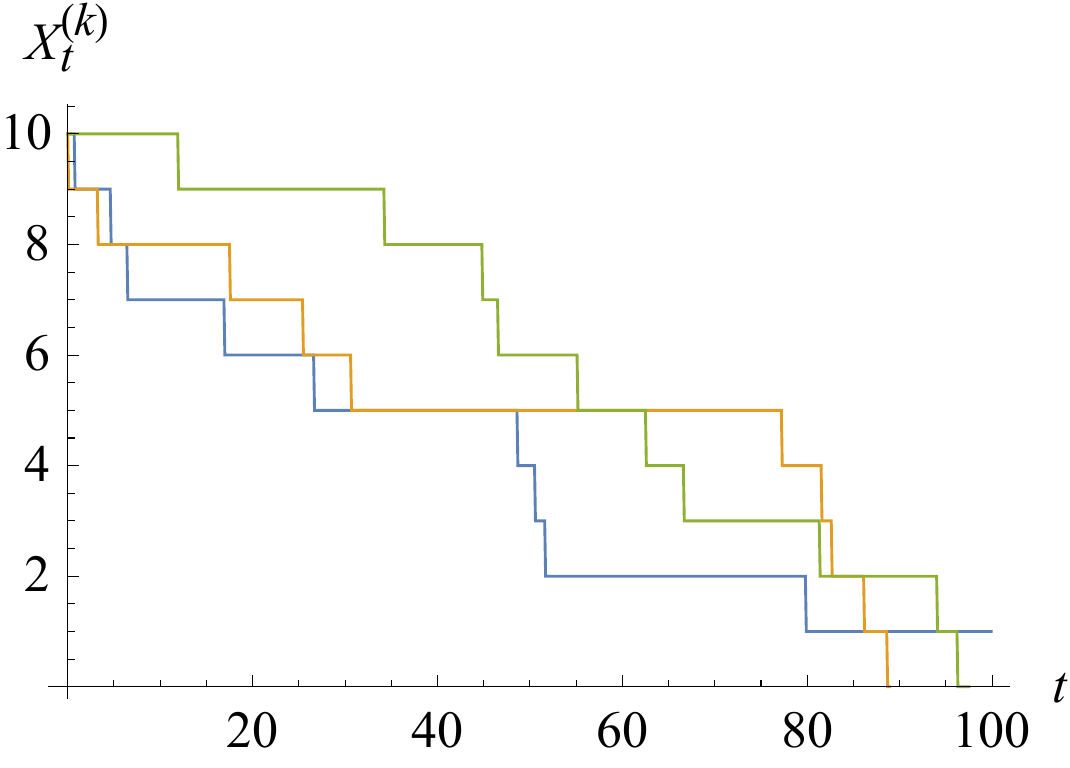}
\caption{Heuristic strategies played against each other: Simulated price trajectories (left window 7a) and inventory levels over time (right window 7b), $K=3$, $N=10$, $T=100$; Example \ref{ex6}.}
\label{fig7}
\end{center}
\end{figure}
%

%
Figure 7a illustrates a scenario in which the heuristic strategy, cf. Algorithm 3.1, is played by multiple firms. We observe that firms undercut each other until a certain price level is obtained. Then firms - one after another - raise their prices up to a certain maximum. This evolution is similar to \blue{the price patterns of Figure 6, cf. Example 5.2}. We observe, that the length of price cycles is longer if less firms are involved.
\blue{The evolution of the inventory levels of all firms is shown in Figure 7b. 
Firms exits the market due to stock-outs ($t=88$ and $t=96$, cf. Figure 7b)}. 
Overall, \blue{we observe that} the heuristic feedback strategy is "reasonably aggressive" but not "destructive" and effectively avoids a decline in price (race to the bottom). 
Note, this character has similarities to equilibrium strategies. 

In case reaction times and inventory levels are observable and firms are rational, best response strategies could be derived using backward induction. This assumption, however, appears not to be a realistic one. Pricing strategies of firms will depend on time and their inventory level, which, in general, is not observable for competitors. Hence, we have competition with asymmetric information. In this setting, equilibrium pricing strategies will be hard to determine. Mutual best response strategies might use observations of offer prices to estimate probability distributions for the competitors' remaining inventory (Hidden Markov Model), \blue{cf. Schlosser, Richly (2018). However, the applicability of such models is again limited} due to the curse of dimensionality.

Finally, our heuristic is a viable and successful strategy to be applied in competitive real-life markets. Moreover, our heuristic is also a suitable strategy to be played as a response against itself.


\subsection{Price Adjustment Costs}

\blue{In the previous sections, we have assumed that price changes are free. Hence, frequent updates can be used to compensate for the lack of market anticipation. 
However, in some markets the number of price updates are limited or may create even (hidden) costs as frequent price adjustments may confuse customers and reduce their trust.
Hence, it can be reasonable to put penalties on price changes in order to implement only necessary price adjustments.
}

\blue{In this context, we present a model with a fixed penalty $z$ for each price adjustment.
To be able to include price adjustment costs in our dynamic programming framework, cf. (4) - (7), we extend the state space by one dimension -- the offer price ${a^ - }$ of the previous period, ${a^ - } \in A$. The costs $z$, $z > 0$, of a price change occur if the price update $a$ differs from the previous price ${a^ - }$. 
To model the case in which no previous price exists, we let ${a^ - }:= 0$.}

\blue{For any market situation $\vec s$ the natural boundary condition for the value function is given by, cf. (4) - (5), 
$n = 0,...,N$, $t = 0,...,T$, ${a^ - } \in A \cup \{ 0\}$,}

\blue{
\begin{equation}
\label{19}
V_T^{}(n,\vec s,{a^ - }) = V_t^{}(0,\vec s,{a^ - }) = 0.
\end{equation}}
\blue{The remaining values ${V_t}(n,\vec s,{a^ - })$ are defined by the HJB equation, $n = 1,...,N$, $t = 0,...,T-1$, ${a^ - } \in A \cup \{ 0\}$,}
\blue{
\[{V_t}(n,\vec s,{a^ - })\]
\begin{equation}
\label{20}
= \mathop {\max }\limits_{a \in A} \left\{ {\sum\limits_{i \ge 0} {{{\tilde P}_t}(i,a|\vec s)}  \cdot \left( {(a - c) \cdot \min (n,i) - n \cdot l - z \cdot {\mathds{1}_{\{ a \ne {a^ - } \wedge {a^ - } > 0\} }} + \delta  \cdot {V_{t + 1}}\left( {{{(n - i)}^ + },\vec s,a} \right)} \right)} \right\},
\end{equation}}

\noindent
\blue{where the indicator function ${\mathds{1}_{\{ a \ne {a^ - } \wedge {a^ - } > 0\} }}$ is 1 if $a \ne {a^ - }$ and ${a^ - } > 0$, it is 0 if $a = {a^ - }$ or ${a^ - } = 0$. The transition of the new state component is straightforward -- the new price $a$ becomes the old price $a^-$ of the next period. Again, the associated pricing strategy is determined by the arg max of (20), $n = 1,...,N$, $t = 0,...,T-1$, ${a^ - } \in A \cup \{ 0\}$.}

\blue{Algorithm 3.1 can still be used in the same way. The numerical complexity is slightly higher but tolerable.
Due to adjustment costs, prices are only updated if opportunity costs (of not updating the price) are sufficiently large. 
The larger $z$ is chosen, the less price updates will be made. 
Note, compared to our previous models, in general, larger delays between two price adjustments (i.e., lower adjustment frequencies) also lead to less price updates, but they sacrifice the advantage of fast reactions to changing markets. Hence, the model above 
allows for fast reactions \textit{and} a reasonable limitation of the number of price updates.}



\section{Evaluation in Practice}

To evaluate \blue{the real-life performance of the presented approach},
we applied our heuristic strategy on Amazon Marketplace. 
Online market platforms such as Amazon
or eBay are highly dynamic as sellers can observe the current 
market situation at any point in time and adjust their prices
instantly. 
This dynamic is hard to manage as pricing decision
requires handling a multitude of dimensions for each competitor 
\blue{(price, quality, seller ratings, etc.).} 

In this experiment, we \blue{have partnered} with a German bookseller. The seller
is among the top 10 sellers for used books on Amazon in
Germany and has an inventory of over 100\,000 distinct books
(ISBN), each with multiple items (1-20). Our seller can decide 
-- to some extent -- on the replenishment of used books
(by adjusting purchase prices). However, supply is limited
and it is not possible to directly reorder specific \blue{(used)} books.
Hence, the challenge is to extract as much profit as possible 
from a given number of books (cf. inventory level) in a
reasonable amount of time.
The pricing strategy of our project partner is characterized 
by a rule-based system that has been developed over
the past years by carefully adjusting rules to lessons learned
from selling books on Amazon. As our project partner has
more than ten years of experience in the market, we consider
his strategy to be effective and accurate. However, market
dynamics are increasingly sophisticated making rule-based
strategies increasingly hard to handle and maintain.

\subsection{Implementation and Architecture}

As our approach is designed to be applied in \blue{practice,
we needed to calibrate} the model, \blue{i.e., holding costs, shipping costs, discount rate, and particularly the sales probabilities, which are based on multiple offer dimensions.
Requested market situations for a certain product from Amazon Marketplace 
%
%
include product-specific features (e.g., Amazon sales rank) as well as merchant-specific features (price, quality, seller ratings, feedback count, etc.). 
The most important offer dimensions are summarized in Table 4.}

\smallskip


\begin{table}[ht]
\centering
\begin{tabular}{l l l}
\toprule
&Offer Dimension     &	Range/Unit	\\ \midrule
\multirow{3}{*}{\rotatebox[origin=c]{90}{\parbox[c]{2.15cm}{\centering Product-Specific\\Parameters}}}
&time	&  seconds\\
&Amazon sales rank of ISBN	&  1 -- 5\,000\,000 by 1\\
&weight	&  gram\\
&original price	& 0.01 -- 500 Euro by 0.01\\
&number of used offers	& 0 -- 20 by 1\\ 
	\cmidrule{1-2}
	\multirow{3}{*}{\rotatebox[origin=c]{90}{\parbox[c]{3.0cm}{\centering Competitor-Specific Parameters}}}
&price	&  0.01 -- 500 Euro by 0.01\\
&condition/quality	&  new -- acceptable (6 levels)\\
&rating	&  0\% -- 100\%\\
&feedback count	&  0 -- 5\,000\,000  by 1\\
&shipping time	&  0 -- 30 days by 1\\
&shipping costs	&  0 -- 10 Euro by 0.01\\
&domestic shipping	& yes / no \\
\bottomrule
\end{tabular}
\caption{\blue{Product- and} merchant-specific offer dimensions.}
\label{tab4}
\end{table}

\blue{Figure 8 illustrates real-life price patterns on the Amazon Marketplace where heterogeneous firms compete for one market. We observe the following typical characteristics: (i) active and passive competitors, (ii) volatile prices with wide price ranges, (iii) entries (cf. Firm 6) and exits of firms (cf. Firm 4), and (iv) Edgeworth-like price cycles. Those characteristics have similarities with our examples and verify the suitability of our models, cf. Section 4 and 5.}


\begin{figure}[ht]
\begin{center}
\includegraphics[width=0.95\textwidth]{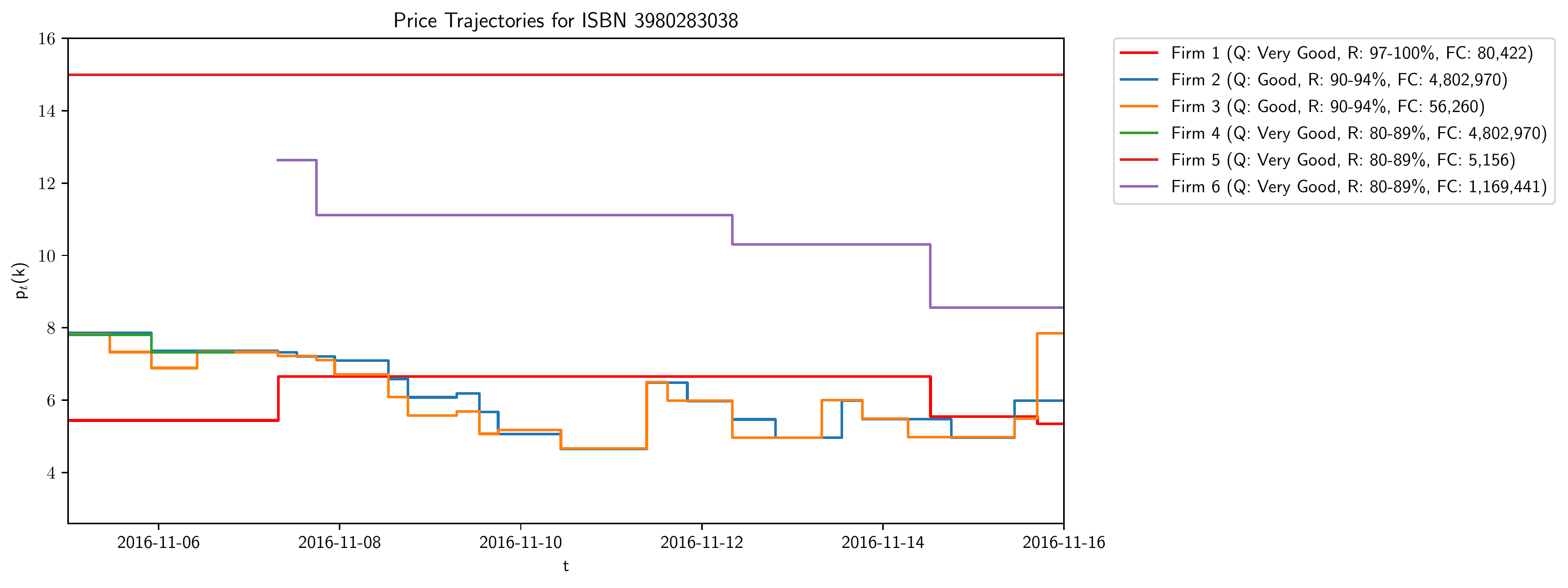}
\caption{\blue{Example of real-life price trajectories from Amazon Marketplace over time for the sale of used books in an oligopoly setting (based on observations in discrete points in time); the legend contains (i) product quality Q, (ii) merchant rating R, and (iii) feedback count FC of all competing firms}.}
\label{fig_prices}
\end{center}
\end{figure}

\begin{figure}[ht]
\begin{center}
\includegraphics[width=\textwidth]{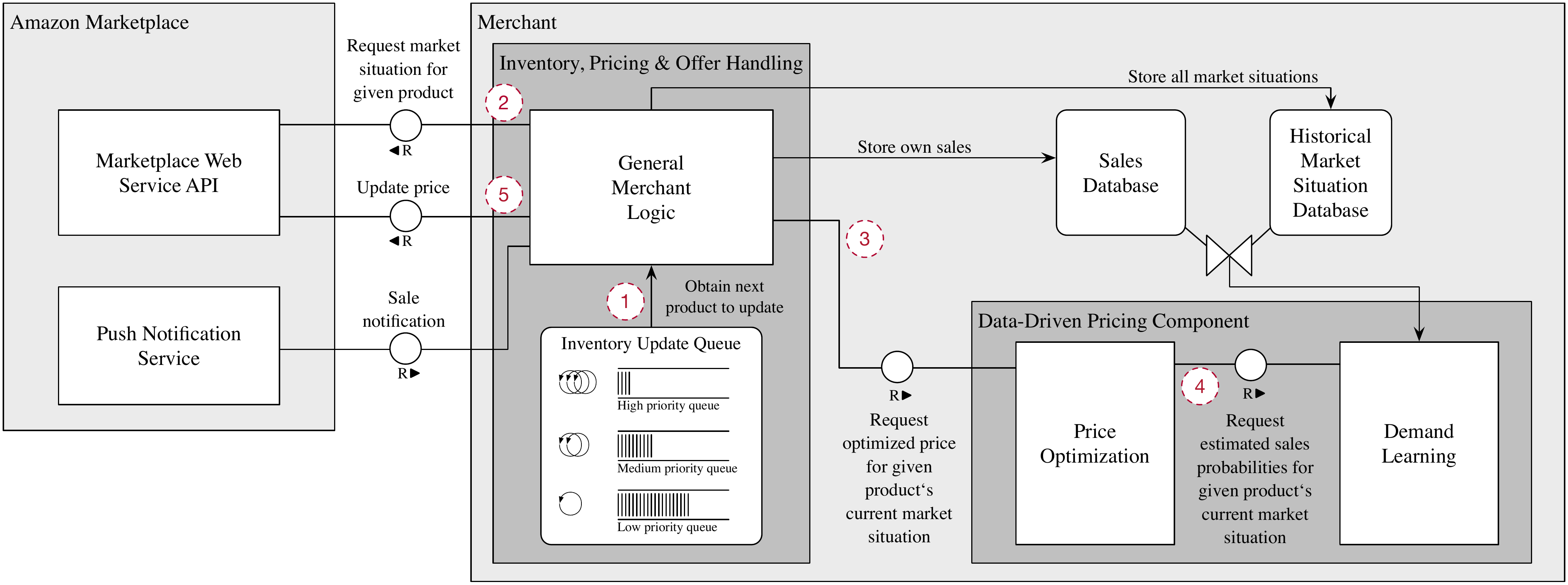}
\caption{\blue{Illustration of the data-driven price update process of a merchant on Amazon Marketplace}.}
\label{fig_process}
\end{center}
\end{figure}

\blue{
\Cref{fig_process} depicts the seller's communication with the Amazon Marketplace and the data flow.
The decision when to update which product is determined by product queues, which separate the seller's products by an adjustable priority (Step 1), cf. Inventory Update Queue.
To update the price of a given product of the inventory, the seller requests the current market situation via the Marketplace Web Service API (Step~2).}
\blue{
The main component 
integrating our heuristic strategy
is the Data-Driven Pricing Component (Step 3).
Based on predicted sales probabilities for a product's market situation (Step 4), cf. Demand Learning, the Price Optimization determines a price and the price update is sent back to Amazon (Step 5).}

\blue{
As the rate with which sellers can request market situations and update prices is limited, products of higher importance are stored in the most frequently updated queue.
On the other hand, products which do not face strong competition or are expected to be accessed less often by customers, are stored in the least frequently updated queue.
The queuing mechanism updates products every 2-12 hours (i.e., 2-10 updates per book per day), which results in $>$20 M observed market situations per month.
Considering on average seven competitors per market situation corresponds to $>$140 M single competitor observations per month. The majority of requests also lead to a price update.}

\bigskip


\blue{
Another used service is the Push Notification Service, cf. \Cref{fig_process}, which informs sellers of products sold on the Marketplace platform.
Sellers can store both data streams (requested market situations and pushed sales) locally in a database.
We join the market data with our partner's price updates, placed orders, and stock data to create the required observations and the corresponding features (cf. Definition 4.1).
Working directly on the raw time-series data provides us with more flexibility, e.g., when regressing only a subset of comparable market situations.
The components regularly train a regression model that allows to predict sales probabilities for any market situation.}


Our estimations of sales probabilities for 
a specific book in a certain time interval are based on market situations, which
\blue{
are characterized by product-specific features
as well as multiple competitor-specific offer dimensions for each present competitor, 
cf. Table 4.}
%
%
\blue{We defined 30 customized features (explanatory variables, cf.~Section 4.1) to describe the relative competitiveness (i.e., sales probabilities) of a single offer in a particular market situation.
We used features similar to those described in Definition \ref{def1} such as, e.g., the price rank of our offer price within the present competitors' prices, etc. Besides price also other offer dimensions were taken into account.}
\blue{
To reduce estimation biases, in an exploration phase, also a certain share of randomized prices updates were used by the seller.}


\subsection{Performance Results}

We tested different demand learning techniques to
quantify how \blue{sales probabilities of specific products are affected by our offer prices in specific market situations.
We decided for a weighted logistic regression approach to estimate probabilities to sell a certain book within a certain time frame at potential offer prices given a particular market situation}.
We calibrated our dynamic programming \blue{model, cf. Section 3,} based on the estimated (conditional) sales probabilities $(\tilde P)$. 
Using our heuristic approach, \blue{we computed
price adjustments for exemplary product offers in specific market situations and checked their plausibility compared to the merchant's strategy and experience.} 
The application of our dynamic pricing strategy works as described in the previous sections, \blue{cf. Algoritm 3.1.
Note, as discussed earlier,} the extension of the dimensionality of the market situation
does not affect the algorithm's complexity.

Finally, we used the calibrated model to determine heuristic pricing strategies to be applied on the Amazon Marketplace.
In our experiment, we used four different discount factors $(\delta^{(1)}>\delta^{(2)}>\delta^{(3)}>\delta^{(4)})$ to vary the strategies' aggressiveness, cf. strategy $S^{(1)}$ -- $S^{(4)}$ in Table 5. $S^{(4)}$ is the most aggressive strategy; \blue{$S^{(1)}$ is the most tame strategy, cf. Remark 4.2 (iii)}.
Over two months, we compared our data-driven strategies with the seller's rule-based benchmark strategy.
To each strategy, we randomly assigned a test group of around 3\,300 books.
\blue{To guarantee a fair comparison of the \textit{quality} of a strategies' pricing decisions, the price adjustment frequencies were the same for the merchant's benchmark strategy and our four data-driven strategies.}


\bigskip

\begin{table}[ht]
\begin{center}
\setlength{\tabcolsep}{2mm}
\begin{tabular}{c c c c c}
\toprule
Strategy   & Test group size	& \% Sold	& Revenue per sale (EUR)	 & Profit per sale (EUR) \\
\midrule
\\[-5pt]
Benchmark	&3\,399	&36.33\quad(100\%)	&6.71 \quad(\hphantom{,}100\%)	&1.90 \quad(\hphantom{,}100\%) \\[2pt]
S$^{(1)}$	&3\,210	&23.24\quad(--36\%)	&8.19 \quad(+22\%)	&3.07 \quad(+61\%) \\
S$^{(2)}$	&3\,339	&31.57\quad(--13\%)	&7.56 \quad(+13\%)	&2.57 \quad(+35\%) \\
S$^{(3)}$	&3\,185	&35.35\quad(--\hphantom{2}3\%)	&7.37 \quad(+10\%)	&2.42 \quad(+27\%) \\
S$^{(4)}$	&3\,155 &37.27\quad(+\hphantom{,}3\%)	&7.00 \quad(+\hphantom{2}4\%)	&2.13 \quad(+12\%) \\
\bottomrule
\end{tabular}
\caption{\blue{Performance comparison} of our data-driven strategies and a \blue{merchant's} rule-based benchmark strategy.
\label{tab5}}
\end{center}
\end{table}
%
%

Table 5 summarizes a comparison of sales, revenues per sale, and profits per sale of the different strategies.
Profits are defined as revenue minus costs, i.e., shipping, Amazon provision, tax (7\%), packing, additional costs (warehouse rent, electricity costs, staff costs), and the average purchase price per item.


As expected, the speed of sales increases and the profitability decreases if our heuristic strategies \blue{are} more aggressive.
Hence, the aggressiveness of \blue{data-driven strategies} can be used to actively control the tradeoff between profitability and speed of sales.
Moreover, strategy $S^{(4)}$ reveals that our approach can sell faster (+3\%) and at the same time more profitable (+12\%) as the seller's benchmark strategy.


\begin{table}[ht]
\centering
\begin{tabular}{cc} 
\toprule
Strategy   & Relative accumulated profit \\
\midrule
\\[-5pt]
Benchmark	&       \hphantom{2}100\% \\[2pt]
S$^{(1)}$	&     +\hphantom{ 1}3\% \\
S$^{(2)}$	&   +\hphantom{ }17\% \\
S$^{(3)}$	&+\hphantom{ }24\% \\
S$^{(4)}$	&   +\hphantom{ }15\% \\
\bottomrule
\end{tabular}
\caption{Comparison of accumulated profits.
\label{tab6}}
\end{table}


In Table 6, we compare the accumulated profits of all strategies.
The relative accumulated profit denotes the quantity "profit per sale (EUR) $\times$ \% of items sold" compared to the corresponding value of the benchmark strategy.
Results show that with our strategy applied, profits can be increased by more than 20\%, cf. S$^{(3)}$.
Moreover, the 
model's discount factor can be used as a management instrument to 
smoothly balance profits, revenues, and the speed of sales according to a seller's needs.

\blue{The individual price patterns of the merchant's benchmark strategy and our data-driven strategy highly depend on (i) the specific article involved, (ii) the number of competitors, (iii) the competitors' offers, (iv) the kind of competitors (commercial, private, strategic) involved, and (v) the price reaction frequencies of competitors. In general, it is advisable not to exclude higher price ranks in advance. Instead, multiple offer dimensions of competitors' offers have to be taken into account. Price adjustment frequencies and reaction times play a prominent role, particularly, when the customer arrival intensity is high.
Further, a successful strategy has to balance two effects: a decline in price by undercutting competitors and moderate price decisions that retain/restore a profitable price level.}

%











\section{Conclusion}

Dynamic pricing under strong competition and incomplete information is a major open problem in revenue management. Practical relevance is enormous, but the problem appears intrinsically hard due to the curse of dimensionality. The challenge is to derive viable heuristic approaches that (i) can be applied in various settings, (ii) allow for data-driven demand estimations, (iii) achieve near-optimal performance results, and (iv) have minimal computation times. 

\blue{We have shown how to compute} dynamic pricing strategies under competition. We have demonstrated that our heuristic approach is applicable even if the number of competitors is large and the competitors' strategies are unknown.

Due to the intractibility of the problem, in general, neither optimal policies nor performance guarantees can be derived.
To measure the performance of our strategy, we computed upper bounds for expected profits, which are obtained by optimal strategies that take advantage of price anticipations. 
We observed that reaction times have a major effect on the performance of dynamic pricing strategies. Moreover, we found that higher adjustment frequencies can easily overcompensate a loss in expected profits due to the lack of price anticipations. Adjusting the price more frequently can be more advantageous than anticipating future market dynamics. Our heuristic approach allows for frequent price adjustments as the computation of prices is efficient and fast.
\blue{Furthermore, in order to choose the number of price adjustments with care without sacrifizing a short reaction time, we present an extended version of the model which allows to take price adjustment costs into account.}
 


Our approach is based on conditional sales probabilities that are estimated from observable market data. 
Market situations are allowed to be \blue{high-dimensional and might also include multiple product features}, such as quality, ratings, shipping time, etc. \blue{Estimated (conditional) sales probabilities make it possible to capture the 
essence of market dynamics as the impact of (even unobservable) market evolutions} on sales probabilities is implicitly included. Thus, it is not necessary to know competitors' strategies explicitly or their reaction times, but we can still take their characteristic behaviour into account. Note, on average, this also includes the impact of exits or entries of new competitors. \blue{Finally, changes of market situations are captured within the framework of regular market observations and price adjustments.}

Using numerical examples that are based on real-life data, we verified the performance of our heuristic pricing strategy. In a real-life experiment, we outperformed the rule-based strategy of an experienced seller by more than 20\%.

Finally, our approach combines key features that are important for real-life applications. First, the approach is applicable if many competitors are involved and offers have \blue{multiple features}. Second, market dynamics do not have to be explicitly known, but they can be indirectly taken into account using data-driven demand estimations. Third, \blue{the computation} of prices is efficient, easy to implement, and allows for frequent price adjustments.








\newpage


\appendix


\section{Additional Figures \& Notation Table}
\label{app}



\begin{figure}[ht]
\begin{center}
\includegraphics[scale=0.32]{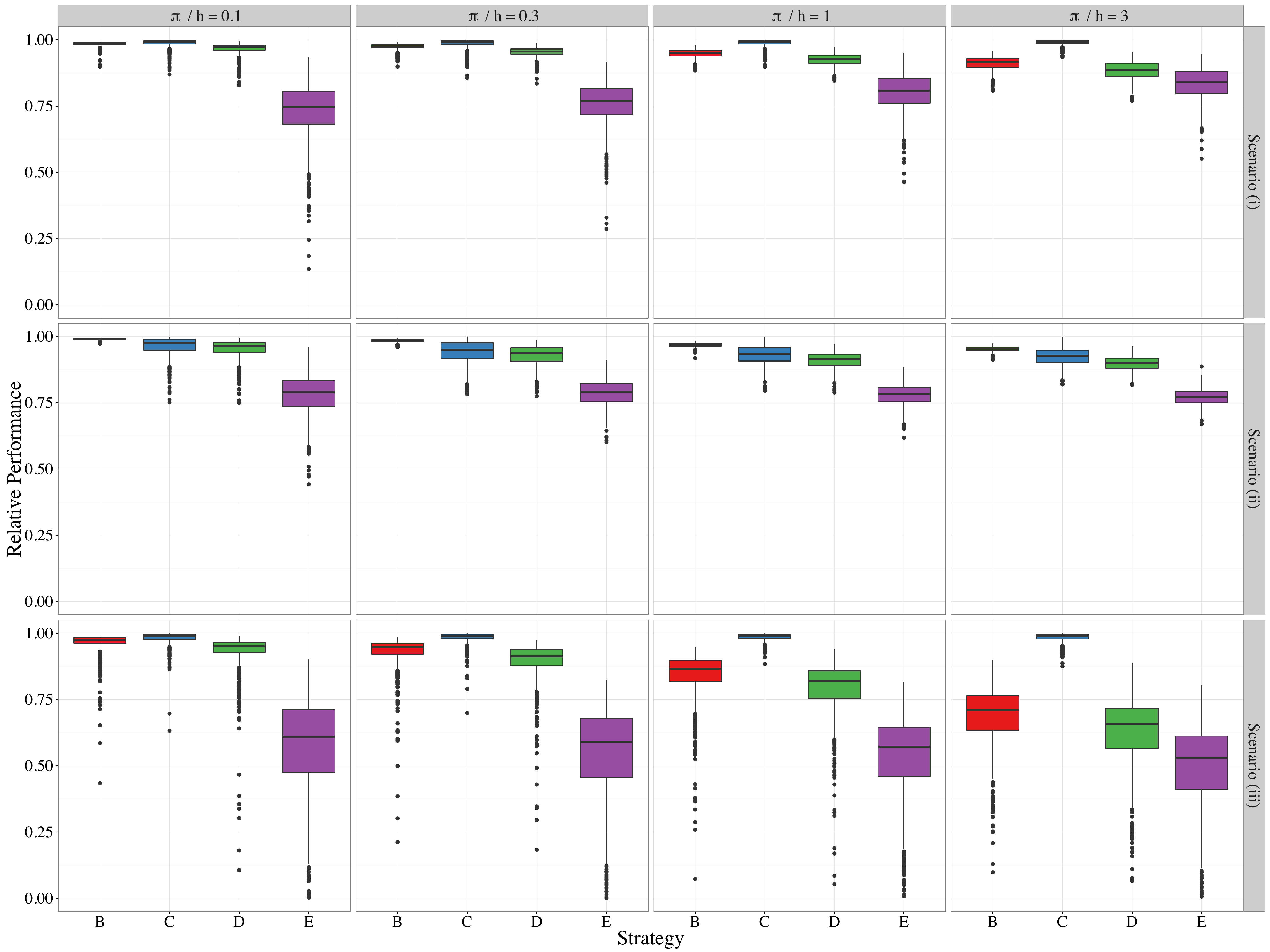} 
\caption{Performance comparison, Boxplot for Table 2; Example \ref{ex4}.}
\label{fig8}
\end{center}
\end{figure}

\vspace{-0.3cm}


\begin{figure}[ht]
\begin{center}
\includegraphics[scale=0.51]{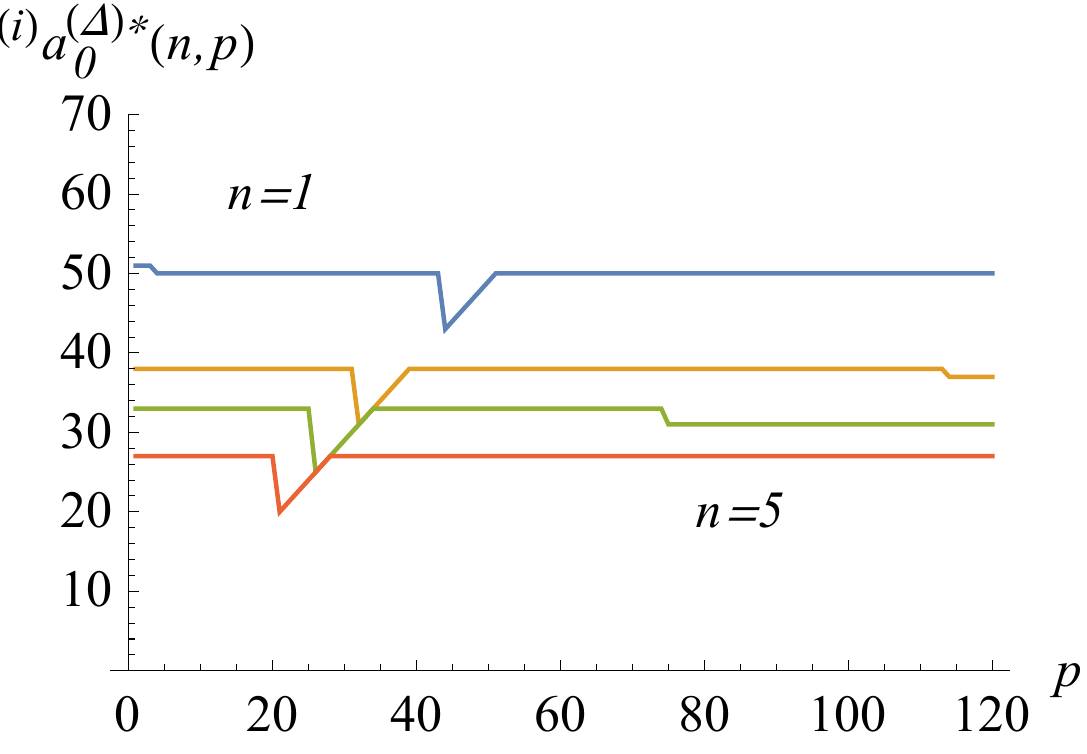}\qquad
\includegraphics[scale=0.51]{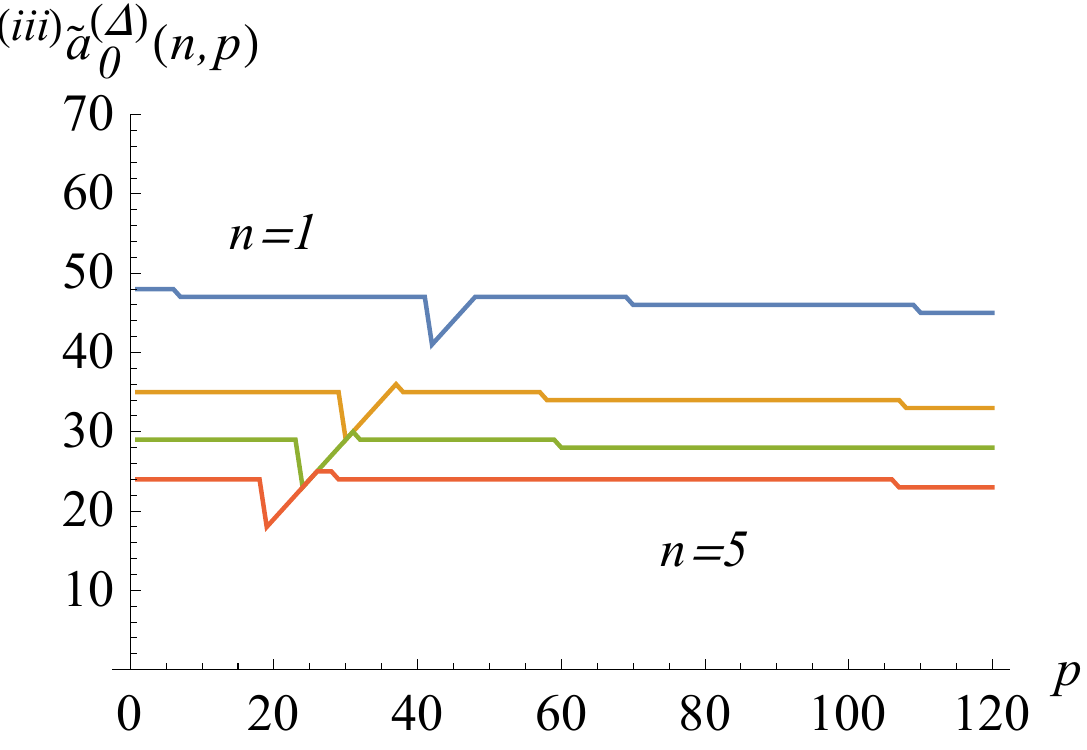}
\caption{Optimal response policy $^{(i)}a_0^{(\Delta )*}(n,p)$ (left window A.11a) and heuristic prices $^{(iii)}\tilde a_0^{(\Delta )}(n,p)$ in case (iii) (right window A.11b) for $n=1, 2, 3, 5$, $T=100$, $\Delta=0.1$; Example \ref{ex5}.}
\label{fig9}
\end{center}
\end{figure}


\begin{figure}[ht]
\begin{center}
\includegraphics[scale=0.51]{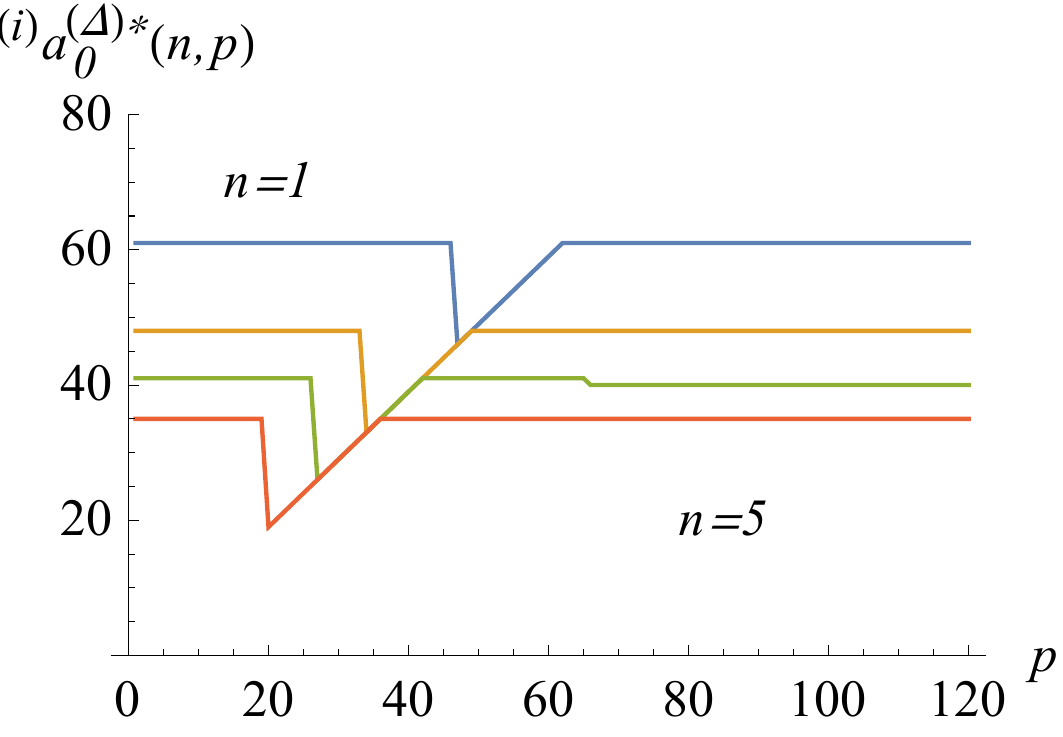}\qquad
\includegraphics[scale=0.51]{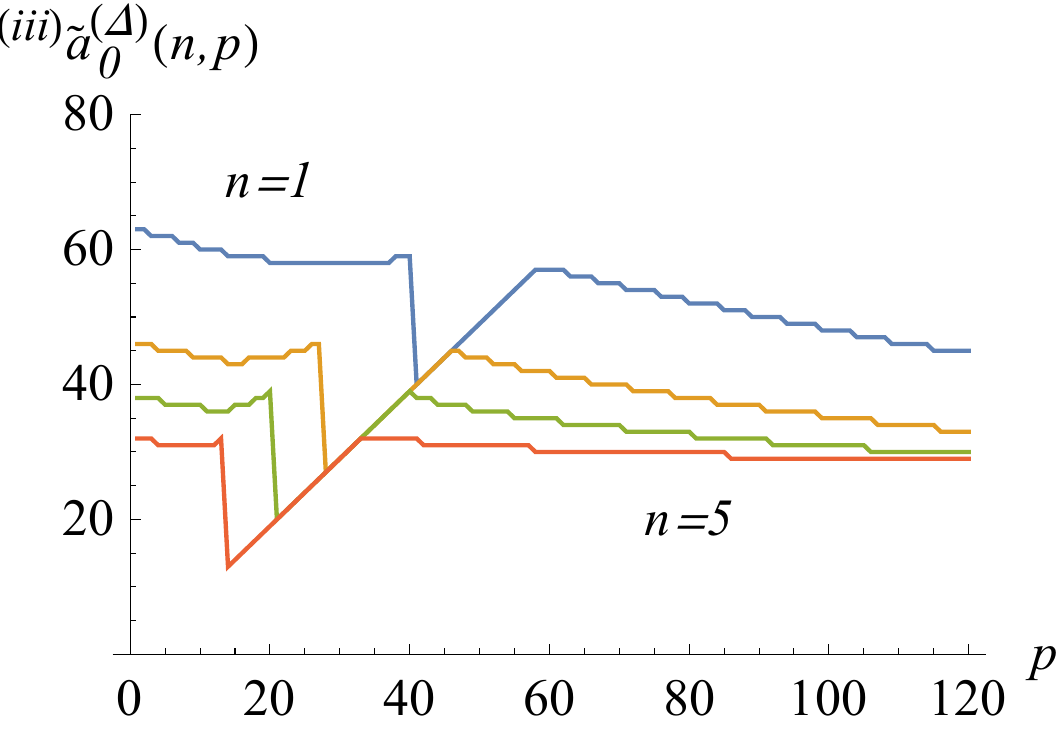}
\caption{Optimal response policy $^{(i)}a_0^{(\Delta )*}(n,p)$ (left window A.12a) and heuristic prices $^{(iii)}\tilde a_0^{(\Delta )}(n,p)$ in case (iii) (right window A.12b) for $n=1, 2, 3, 5$, $T=100$, $\Delta=0.9$; Example \ref{ex5}.}
\label{fig10}
\end{center}
\end{figure}


\begin{table}[ht]
\begin{center}
\setlength{\tabcolsep}{3mm}
\begin{tabular}{cccc}
\\ \hline
$t$			&time / period				&	$a$				&offer price \\
$N$			&initial number of items to sell	&	$V$, $\bar V$ 		&value functions   \\
$X$			&number of items left to sell	&	$\vec x$, $\vec s$ 	&market situation  \\
$c$			&shipping costs				&	$\lambda$ 		&sales intensity  \\
$\delta$   		&discount factor				&	$P$    			&sales probability  \\
$l$			&holding costs				&	$\tilde P$			&conditional sales probability  \\
$h$			&length of a subperiod		&	$r$				&sales rank  \\
$G$			&future profits				&	$\vec p$			&competitors' prices  \\
$K$			&number of competitors		&	$\pi$				&reaction probability  \\
$A$			&set of admissible prices		&	$\Delta$			&reaction time \\
$T$			&time horizon				&	$\vec \beta$		&regression coefficients  \\
$z$			&\blue{costs of a price change}		&	$a^{-}$		&\blue{previous price}  \\ \hline
\end{tabular}
\caption{List of variables and parameters.
\label{tab7}}
\end{center}
\end{table}

\end{document}